\documentclass[journal,twoside]{IEEEtran}
\usepackage{csquotes}
\usepackage{tkz-graph}
\usepackage{enumitem}
\setlist{leftmargin=6mm}
\usetikzlibrary{calc}
\tikzset{me/.style={to path={
\pgfextra{%
 \pgfmathsetmacro{\startf}{-(#1-1)/2}
 \pgfmathsetmacro{\endf}{-\startf}
 \pgfmathsetmacro{\stepf}{\startf+1}}
 \ifnum 1=#1 -- (\tikztotarget)  \else
     let \p{mid}=($(\tikztostart)!0.5!(\tikztotarget)$)
         in
\foreach \i in {\startf,\stepf,...,\endf}
    {%
     (\tikztostart) .. controls ($ (\p{mid})!\i*6pt!90:(\tikztotarget) $) .. (\tikztotarget)
      }
      \fi
     \tikztonodes
}}}

\usepackage{centernot}
\newcommand{\argmin}{\operatornamewithlimits{argmin}}

\usepackage{amsfonts,amssymb}
\usepackage{color}
\usepackage{pifont}
\usepackage{multirow}
\usepackage{amsmath}
\usepackage{url}
\usepackage{graphicx}
\usepackage{ntheorem}
 \usepackage{mathrsfs}
\usepackage{cite}
  \usepackage{cases}
 \usepackage{pbox}

\usepackage[font=normalsize,labelfont=normalsize]{caption}
\def\hlinewd#1{%
\noalign{\ifnum0=`}\fi\hrule \@height #1 %
\futurelet\reserved@a\@xhline}

\newcommand{\given} {{\mathlarger{\boldsymbol\mid}}\hspace{1pt}}
\newcommand{\bgiven} {{\mathlarger{\mathlarger{\mathlarger{\mathlarger{\mid}}}}}}

\newcommand{\bcup} {\hspace{2pt} \mathlarger{\cup}
\hspace{2pt}}

 \usepackage{epsfig}
 \newfont{\secfnta}{ptmb8t at 12pt}

\usepackage{makecell}
\usepackage[top=.65in,bottom=.65in,left=.5in,right=.5in]{geometry}
\usepackage{relsize}

\newcommand{\bcap} {\hspace{2pt} \mathlarger{\cap}
\hspace{2pt}}

    \makeatletter
    \def\@listi{\leftmargin\leftmargini
        \parsep 1\p@ \@plus0\p@ \@minus\p@
        \topsep 2\p@   \@plus0\p@ \@minus\p@
        \itemsep1\p@ \@plus0\p@ \@minus\p@}
    \let\@listI\@listi\@listi
    \makeatother

\newcommand*\mcapinn[2]{\vcenter{\hbox{$\mathsurround=0pt
\ifx\displaystyle#1\textstyle\else#1\fi\bigcap$}}}

\newcommand*\mcupinn[2]{\vcenter{\hbox{$
\bigcup$}}}

\newcommand{\bP}[1]{{\mathbb{P}}\left[{#1}\right]}

\newtheorem{fact}{Fact}
\newtheorem{lem}{Lemma}
\newtheorem{cla}{Claim}
\newtheorem{defn}{Definition}
\newtheorem{thm}{Theorem}

\newtheorem{rem}{Remark}

\begin{document}

\title{On Connectivity and Robustness in Random Intersection Graphs}


\author{Jun~Zhao,~\IEEEmembership{Student Member,~IEEE,}
        Osman~Ya\u{g}an,~\IEEEmembership{Member,~IEEE,}
        and~Virgil~Gligor,~\IEEEmembership{Senior Member,~IEEE}
        \thanks{The
material in this paper was presented in part at the 2014 IEEE Conference on Decision and Control (CDC), Los Angeles, CA, USA \cite{ZhaoCDC}.

J. Zhao was, and O. Ya\u{g}an and V. Gligor are with the Cybersecurity Lab (CyLab) and the Department of Electrical and Computer Engineering,
Carnegie Mellon University, Pittsburgh, PA 15213, USA. (Emails: junzhao@alumni.cmu.edu, oyagan@ece.cmu.edu, gligor@cmu.edu). J. Zhao is now with Arizona State University.

This research was supported in part by CyLab, and Department of Electrical \& Computer Engineering
at Carnegie Mellon University, and also by
Grants CCF-0424422 and CNS-0831440 from the National Science Foundation to the Berkeley TRUST STC.

}
}

\markboth{IEEE Transactions on Automatic Control}{IEEE Transactions on Automatic Control}

\maketitle


\begin{abstract}

Random intersection graphs have received much attention recently and been used in a wide range of applications ranging
from key predistribution in wireless sensor
 networks to modeling social networks. For these graphs,
each node is equipped with a set
of objects in a random manner, and two nodes have an
undirected edge in between if they
have at least one object in common.
 In this paper, we investigate connectivity and robustness in a {\em general random
intersection graph} model. Specifically, we establish sharp
asymptotic zero--one laws for
 $k$-connectivity and $k$-robustness, as well as the
asymptotically exact probability of $k$-connectivity,
 for any positive integer $k$.
The $k$-connectivity property quantifies how resilient is the
connectivity of a graph against node or edge failures, while $k$-robustness measures the effectiveness of {\em local-information-based}
 consensus algorithms (that do not use global graph topology
information) in the presence
of adversarial nodes.
  In addition to presenting the results under the general random
intersection graph model, we consider two special cases of the
general model, a {\em binomial} random intersection graph and a {\em
uniform} random intersection graph, which both have numerous applications as well.
For these two specialized graphs, our results on asymptotically
exact probabilities of $k$-connectivity and asymptotic zero--one laws for $k$-robustness are also novel in the literature.

\end{abstract}


\section{Introduction}

\subsection{Background} \label{sec:GraphModel}

Since random intersection graphs were introduced by Singer-Cohen
\cite{RIGThesis}, different classes of these graphs have received considerable attention \cite{Rybarczyk,yagan,virgil,2013arXiv1301.0466R,zz,r1,ryb3,bloznelis2013}
recently.
In these graphs, each node is assigned a set
of objects selected by some random mechanism. An undirected edge exists between any two nodes that
have at least one object in common.
Random intersection graphs have been used in modeling and analyzing real-world networks in a wide variety of applications. Examples include
secure wireless sensor networks  \cite{Rybarczyk,yagan,virgil}, social networks \cite{ryb3,zz,2013arXiv1301.0466R},
 classification analysis \cite{Models},
 and cryptanalysis \cite{herdingRKG}.
 Several properties such as clustering \cite{bloznelis2013},
component evolution \cite{Rybarczyk} and degree
distribution \cite{Models} have been analyzed for different classes of
random intersection graphs.

The graph model  in this paper, hereafter referred to as a {\em
general random intersection graph}, represents a
 generalization of the random intersection graphs studied by
Bloznelis {\em et al.} \cite{bloznelis2013,Rybarczyk}, and is defined
on a node set $\mathcal {V}_n = \{v_1,
v_2, \ldots, v_n \}$ as follows. Each node $v_i$ ($i=1,2,\ldots,n$)
is assigned an object set $S_i$ from an object pool $\mathcal {P}_n$ \label{poolref}
 consisting of $P_n$ distinct objects, where $P_n$ is a function of
$n$. Each object set $S_i$ is constructed from the following two steps: First, the size of $S_i$, $|S_i|$, is determined
according to some probability distribution $\mathcal {D}_n:\{1,
2,\ldots, P_n\} \to [0,1]$. Of course, we have $\sum_{x = 1}^{P_n}
\mathbb{P}[|S_i| = x] = 1$, with $\mathbb{P}[A]$ denoting the
probability that event $A$ occurs. Next, $S_i$ is formed by
selecting $|S_i|$ distinct objects uniformly at random from the
object pool $\mathcal {P}_n$. In other words, conditioning on $|S_i| =
s_i$, set $S_i$ is chosen uniformly among all $s_i$-size subsets of
$\mathcal {P}_n$. This process is repeated independently for all
object sets $S_1, \ldots, S_n$. Finally, an undirected edge is
assigned between two nodes if and only if their corresponding object
sets have at least one object in common; namely, distinct nodes
$v_i$ and $v_j$ have an undirected edge in between if and only if $S_i \bcap
S_j \neq \emptyset$. The graph defined through this adjacency notion
is denoted by $G(n,P_n,\mathcal {D}_n)$.

A specific case of the general model $G(n,P_n,\mathcal {D}_n)$,
known as the \emph{binomial} random intersection graph, has
been widely explored to date [9]--[14].
Under this
model, each object set $S_i$ is constructed by a Bernoulli-like
mechanism; i.e., by adding each object to $S_i$ independently with
probability $p_n$. Like integer $P_n$, probability $p_n$ is also a function of $n$. The term \lq\lq
binomial" accounts for the fact that $|S_i|$ now follows a binomial
distribution with $P_n$ as the number of trials and $p_n$ as the
success probability in each trial. We denote the binomial random
intersection graph by $G_b(n,P_n,p_n)$, where the subscript ``b'' stands for ``binomial''.

Another well-known special case of the general model $G(n,P_n,\mathcal
{D}_n)$ is the \emph{uniform} random intersection graph
\cite{r1,yagan,ryb3,virgil}. Under the uniform model, the
probability distribution $\mathcal {D}_n$ concentrates on a single
integer $K_n$, where $1\leq K_n \leq P_n$; i.e., for each node
$v_i$, the object set size $|S_i|$ equals $K_n$ with probability
$1$. Note that $P_n$ and $K_n$ are both integer functions of $n$. We denote by $G_u(n,P_n,K_n)$ the uniform random intersection
graph, with the subscript ``u'' meaning ``uniform''. \vspace{-5pt}

\subsection{Applications of Random Intersection Graphs}

A concrete example for the application of random intersection graphs can be given in the context of secure wireless sensor networks. As explained in detail in numerous other places \cite{Rybarczyk,ryb3,yagan,yagan_onoff,yagan2015zero,ZhaoYaganGligor}, the uniform random intersection graph model $G_u(n,P_n,K_n)$
is induced naturally by the Eschenauer--Gligor random key predistribution
scheme \cite{virgil}, which is a typical solution to ensure secure
communications in wireless sensor networks. In particular, let the set of $n$ nodes in graph
$G_u(n,P_n,K_n)$ stand for the $n$ sensors in the wireless network. Also, let the object pool $\mathcal{P}_n$ (with size $P_n$)
represent the set of cryptographic keys available to the network and let $K_n$ be the number of keys assigned to each sensor
(selected uniformly at random from the key pool $\mathcal{P}_n$). Then, the edges in $G_u(n,P_n,K_n)$ represent pairs of sensors that share at least one cryptographic key and thus that can
{securely} communicate over existing wireless links in the Eschenauer--Gligor scheme. In the above application, objects that nodes have are cryptographic keys, so uniform random intersection graphs are also referred to as random key graphs \cite{yagan,yagan2015zero,yagan_onoff,ZhaoYaganGligor}.

In the secure sensor network area, the general random intersection graph model captures the differences that may exist among the number of keys assigned to each sensor. These differences appear for a variety of reasons including
(a) the number may vary from sensor to sensor in a heterogeneous sensor network due to differences in
the sizes of sensor memories \cite{Rybarczyk};
(b) the number may decrease due to the revocation of compromised nodes and keys  \cite{ChenGligorPerrigMuralidharanTDSC2005}; and
(c) the number may increase due to the establishment of path keys, where new
keys are generated and distributed to participating sensors after deployment \cite{virgil}.

Random intersection graphs can also be used to model social networks, where a node represents an individual, and an object could be an hobby of individuals, a book being read, or a movie being watched, etc. \cite{bloznelis2013,2013arXiv1301.0466R,Bloznelis201494,ZhaoYaganGligor}.
Then a link between two individuals characterizes a common-interest relation;
e.g., two individuals have a connection if they have a common hobby, read the same book, or watch the same movie.
In this setting, binomial/uniform/general random intersection graphs represent common-interest
networks where the sets of interests that individuals have are constructed in different ways.
Specifically, in binomial random intersection graphs, each interest is attached to each person
independently with the same probability; in uniform random intersection graphs, all individuals
have the same number of interests;
and general random intersection graphs provide general possibilities for assigning
individuals' interest sets; e.g., without probability or number-of-interest restrictions.

\subsection{Problem Formulation}
\label{sec:ConnectivityandRobustness}


We now introduce the graph properties that we are interested in. First, $k$-connectivity is formally defined as follows.

\begin{defn}[$k$-Connectivity \cite{bollobas}]
A graph is said to be $k$-connected if each pair of nodes has at least $k$
internally node-disjoint path(s) in between, where two paths are internally node-disjoint if except the source and destination, the intermediate nodes are different. Equivalently, by Menger's theorem, a graph is $k$-connected if
it cannot be disconnected by deleting at most $(k-1)$ nodes or edges, where a graph is connected if there exists at least a path of edges between any two nodes.
\end{defn}
Clearly, $k$-connectivity quantifies well-established measures of strength.
For instance, it captures the resiliency of graphs against node or edge failures.
It also captures the resiliency of consensus protocols in the presence of
$h$ adversarial nodes in a graph with node size greater than $3h$;
i.e., a necessary and sufficient condition is that the graph is $(2h+1)$-connected  \cite{Dolev:1981:BGS:891722}.

Many graph algorithms rely on sufficient connectivity; e.g,  algorithms to
achieve consensus~\cite{6425841,add-ref-2,zhang2012robustness}.
However, these algorithms typically assume that nodes have full knowledge of the graph topology,
which is often impractical~\cite{6425841}.
To account for the lack of full topology knowledge in the general case, Zhang and
Sundaram introduce the notion of {\em graph robustness}~\cite{6425841},
which has received  much attention recently
\cite{6481629,leblanc2013resilient,zhang2012robustness,zhang2012simple,7061412}.
Formally, $k$-robustness is   defined as follows.
\begin{defn}[$k$-Robustness]
A graph with a node set $\mathcal {V}$ is $k$-robust
 if at least one of (a) and (b) below hold for every pair of non-empty, disjoint subsets $A$ and $B$ of $\mathcal {V}$: (a) there exists
at least a node $v_a \in A$ such that
  $v_a$ has no less than $k$ neighbors inside $\mathcal {V}\setminus
  A$; and (b) there exists at least a node $v_b \in B$ such that
  $v_b$ has no less than $k$ neighbors inside $\mathcal {V}\setminus B$.
  \end{defn}
Zhang and Sundaram \cite{6425841} show that
when nodes have local topology knowledge,
consensus can still be reached in a sufficiently robust graph in the presence of adversarial nodes, but not in a sufficiently connected and insufficiently robust graph.

Graph robustness provides a different notion of strength than $k$-connectivity.
That is, it quantifies the effectiveness and resiliency of local-information-based consensus algorithms
in the presence of adversarial nodes. We detail the application of robustness to consensus in the next subsection.
 Robustness also has broad relevance in graph processes beyond consensus;
e.g., robustness plays a key role in information cascades \cite{6425841}.

\subsection{Application of Robustness to Consensus}

To study consensus in a graph, we consider that all nodes are synchronous and the time is divided into different time slots. From one time slot to the next time slot, each node updates its value. Let $x_i[t]$ denote the value of node $v_i$ at time slot $t$ for $t=0,1,\ldots$. We first suppose all nodes are benign. Then consensus means
 $\lim_{t \to \infty} |x_i[t] - x_j[t]| = 0 $ for each pair of nodes $v_i$ and $v_j$.
 The updating process of each node's value is as follows. With $V_i$ denoting the neighborhood set of each node $v_i$, from time slot $t$ to $t+1$, $v_i$ updates its value $x_i [t]$ to $x_i[t+1]$ by incorporating node $v_j$'s value $x_j [t]$ that $v_j$ sends to $v_i$, for $v_j \in V_i$; i.e.,  there is a function $f_i(\cdot)$ such that
\begin{align}
x_i[t+1] & = f_i\big( \big\{ x_j[t] \ \big| \ v_j \in V_i \bcup \{ v_i \} \big\}\big) .  \nonumber
\end{align}
Now we consider the case where there might exist adversarial nodes, i.e., nodes that are not benign. A node $v_i$ is said to be benign if it sends
$x_i [t]$ to all of its neighbors and applies $f_i(\cdot)$ at every time
slot $t$, and is called adversarial otherwise. In the presence of adversarial nodes, consensus means $\lim_{t \to \infty} |x_i[t] - x_j[t]| = 0 $ for each pair of \emph{benign} nodes $v_i$ and $v_j$.

Under the adversary model that the total number of adversarial node(s) in
the graph is upper bounded by $h$, then consensus can be achieved if and only if the graph is $(2h+1)$-connected, given the graph has more than $3h$ nodes \cite{Dolev:1981:BGS:891722}. However, the algorithms often assume that all nodes know the entire network topology \cite{6425841}.
Suppose each node does not know the entire network topology and only knows the number of adversarial
nodes in its neighborhood, Zhang and Sundaram \cite{6425841} show the usefulness of robustness in studying consensus. Specifically, under the adversary model that each benign node has at most $h$ adversarial node(s) as neighbors, then consensus can be achieved if the graph is $(2h+1)$-robust \cite{6425841}. With the above, we can use consensus dynamics
to motivate the study of both connectivity and robustness,
where connectivity (resp., robustness) is applicable to the case where each node knows the global (resp., local) network topology.

\subsection{Related Work} \label{related}

For connectivity (i.e., $k$-connectivity with $k=1$) in binomial
random intersection graph $G_b(n,P_n,p_n)$, Rybarczyk establishes
the exact probability \cite{2013arXiv1301.0466R} and a zero--one law
\cite{zz,2013arXiv1301.0466R}. 
 She
further shows a zero--one law for $k$-connectivity
\cite{zz,2013arXiv1301.0466R}. Our
Theorem \ref{thm:rig} provides not only a zero--one law,
 but also the exact probability to understand $k$-connectivity precisely.

 For connectivity in
uniform random intersection graph $G_u(n,P_n,K_n)$, Rybarczyk
\cite{ryb3} derives the exact probability and a zero--one law, while
Blackburn and Gerke \cite{r1}, and Ya\u{g}an and Makowski
\cite{yagan}
 also obtain zero--one laws. Rybarczyk \cite{zz}
implicitly shows a zero--one law for $k$-connectivity in
$G_u(n,P_n,K_n)$. Our Theorem \ref{thm:urig} also gives a zero--one law. In addition, it gives the exact probability to
provide an accurate understanding of $k$-connectivity.

For general random intersection graph $G(n,P_n,\mathcal {D}_n)$,
Godehardt and Jaworski \cite{Models} investigate its degree
distribution and Bloznelis \emph{et al.} \cite{Rybarczyk} explore
its component evolution.
 Recently,
Ya\u{g}an \cite{yagan2015zero} obtains a zero--one law for connectivity.


Since asymptotic probability results of $k$-connectivity in random graphs are often established by first showing the corresponding results for the property of having minimum degree at least $k$, and then proving the probability of having minimum degree at least $k$ yet not being $k$-connected converges to zero asymptotically, all the above references on $k$-connectivity (resp., connectivity) also establish the corresponding results for the property of having minimum degree at least $k$ (resp., $1$).

To date, there have not been results on
($k$-)robustness of random intersection graphs reported  by others. As noted
in Lemma \ref{er_robust}, Zhang and Sundaram \cite{6425841} present
a zero--one law for $k$-robustness in an Erd\H{o}s--R\'{e}nyi graph.

For random intersection graphs in this paper, two nodes have an edge in between if their object sets share at least one object. A natural variant is to define graphs with edges only between nodes which have at least $s$ objects in common (instead of just $1$) for some positive integer $s$. Recent researches \cite{Bloznelis201494,ANALCO} investigate $k$-connectivity in graphs under this definition.


\subsection{Contributions and Organization}

With the above notions of $k$-connectivity and $k$-robustness in mind,
a natural questions to ask is when will random intersection graphs become
$k$-connected or $k$-robust? We answer this question
 and summarize the key contributions as follows:
\begin{enumerate}
  \item [i)] We derive sharp zero--one laws and asymptotically
  exact probabilities for
$k$-connectivity in general random intersection graphs.
  \item [ii)] We establish sharp zero--one laws for
$k$-robustness in general random intersection graphs.
 \item [iii)] For the two specific instances of
the general graph model, a binomial random intersection graph and a
uniform random intersection graph, we provide the first results on   asymptotically exact probabilities of $k$-connectivity and zero--one
laws for $k$-robustness.
\end{enumerate}

This paper extends the conference version \cite{ZhaoCDC} in several ways:
\begin{enumerate}
  \item [i)] We strengthen the known results on binomial/uniform/general random intersection graphs. Specifically, Theorems 1--6 in this paper eliminate the condition $|\alpha_n| = o(\ln n)$ in \cite[Theorems 1--6]{ZhaoCDC}.
   \item [ii)] For $k$-connectivity of a
uniform random intersection graph, we provide a complete proof in Section \ref{prf:thm:urig}.
Note that this result serves as the building block for all other results.
  \item [iii)] We enhance numerical experiments to better confirm the theoretical results; see Section \ref{sec:expe}.
    \item [iv)] We discuss the parameter conditions of the theorems in detail; see Section \ref{ParameterConditions}.
  \item [v)] We compare our results of binomial/uniform/general random intersection graphs with those of Erd\H{o}s--R\'{e}nyi graphs; see the last paragraph of Section \ref{sec:main:res}.
  \end{enumerate}

The rest of the paper is organized as follows. Section \ref{sec:main:res} presents the main results as
Theorems  \ref{thm:rig}--\ref{thm:grig:rb}. To improve the readability of the paper, we defer the proofs of the theorems to the end of the paper. We provide
numerical experiments in Section \ref{sec:expe}. Afterwards, we introduce some auxiliary lemmas in
Section \ref{sec:factlem}, before establishing the main results in
Sections \ref{sec:thmprf:kcon}, \ref{prf:thm:urig} and \ref{sec:thmprf:krb}. Section
\ref{sec:prf:fact:lem} details the proofs of the lemmas. Finally, we conclude the paper Section \ref{sec:Conclusion}.

\section{The Results} \label{sec:main:res}

Our main results are presented in Theorems \ref{thm:rig}--\ref{thm:grig:rb} below. We defer the proofs of all
theorems to Sections \ref{sec:thmprf:kcon}--\ref{sec:thmprf:krb}.
Throughout the paper, $k$ is a positive integer and does not scale
with $n$; and $e$ is the base of the natural logarithm function,
$\ln$. All limits are understood with $n\to \infty$. We use the
standard Landau asymptotic notation $o(\cdot), O(\cdot),
\omega(\cdot), \Omega(\cdot),\Theta(\cdot)$ and $ \sim$; see \cite[Page 2-Footnote 1]{ZhaoYaganGligor}. In
particular, for two positive sequences $f_n$ and $g_n$, the relation $f_n \sim
g_n$ signifies $\lim_{n \to
  \infty} ( {f_n}/{g_n})=1$.
   For a random variable $X$, the\vspace{1pt} terms $\mathbb{E}[X]$
and $\text{Var}[X]$ stand for its expected value and variance,
respectively.

As noted in Section \ref{sec:GraphModel}, we denote a binomial (resp., uniform)  random intersection graph by $G_b(n,P_n,p_n)$ (resp., $G_u(n,P_n,K_n)$). Clearly, $G_b(n,P_n,0)$ (resp., $G_u(n,P_n,0)$) is an empty graph, while $G_b(n,P_n,1)$ (resp., $G_u(n,P_n,P_n)$) being a complete graph is $k$-connected for $n\geq k+1$ and is $k$-robust for $n\geq 2k$. Then for each $n\geq 2k$, with $P_n$ fixed and $p_n$ increasing from $0$ to $1$ (resp., $K_n$ increasing from $0$ to $P_n$),
the probabilities of $k$-connectivity and $k$-robustness of $G_b(n,P_n,p_n)$ (resp., $G_u(n,P_n,K_n)$)  increase from $0$ to $1$.
In addition, for random graphs, results are often obtained in the asymptotic sense since the analysis
becomes intractable in the finite regime \cite{citeulike:4012374,erdos61conn,RIGThesis,YaganThesis,yagan_onoff}.

Given the above, it is natural to anticipate that our results are  presented in the form of zero--one laws, where a zero--one law means that the probability of a graph having a certain property asymptotically converges to $0$ under some conditions and to $1$ under some other conditions. Moreover, it is useful to have a complete picture by obtaining the asymptotically exact probability result \cite{YaganThesis}. For binomial/uniform/general random intersection graphs, we derive asymptotically exact probabilities for
$k$-connectivity in Theorems  \ref{thm:rig}--\ref{thm:grig}, and zero--one laws for $k$-robustness in Theorems  \ref{thm:rig:rb}--\ref{thm:grig:rb}. A future work is to establish asymptotically exact probabilities for
$k$-robustness.

Noting that for any graph/network, $k$-connectivity implies that the minimum node degree is at least $k$ \cite{citeulike:4012374}, we often present results for the property of minimum node degree being at least $k$ together with $k$-connectivity results.

\setlength{\belowdisplayskip}{2.6pt plus 1.0pt minus 1.0pt}%
\setlength{\belowdisplayshortskip}{1.0pt plus 2pt}
 \setlength{\abovedisplayskip}{2.6pt plus 1.0pt minus 1.0pt}%
\setlength{\abovedisplayshortskip}{1.0pt plus 2pt}

\subsection{Asymptotically Exact Probabilities for
$k$-Connectivity and the Property of Minimum Node Degree Being at Least $k$} \label{sec:kconasy}

\subsubsection{$k$-Connectivity and Minimum Node Degree in   Binomial Random Intersection
Graphs}~

For a binomial random intersection graph,
Theorem \ref{thm:rig}  below shows asymptotically exact probabilities for
$k$-connectivity and the property of minimum node degree being at least $k$.

\begin{thm} \label{thm:rig} For a binomial random intersection graph
$G_b(n,P_n,p_n)$, with a sequence $\alpha_n$ for all $n $ defined
through\vspace{2.00000pt}
\begin{align}
  {p_n}^2 P_n & =
 \frac{\ln  n + {(k-1)} \ln \ln n + {\alpha_n}}{n},  \label{thm:rig:pe}\vspace{2.00000pt}
\end{align}
if $P_n = \omega \big(n(\ln n)^5\big)$,\vspace{2.00000pt}
\begin{align}
 &  \hspace{-25pt} \lim_{n \to \infty}\mathbb{P}
\big[\hspace{2pt}\text{Graph }G_b(n,P_n,p_n)\text{
is $k$-connected}.\hspace{2pt}\big] \nonumber
 \end{align}
 \vspace{-1pt}
 \begin{subnumcases}{ \quad =}
 0, &\hspace{-11.5pt}\text{ if $\lim_{n \to \infty}{\alpha_n}
=-\infty$}, \vspace{2.00000pt}\label{bin-kon-0} \\  1, &\hspace{-11.5pt}\text{ if $\lim_{n \to \infty}{\alpha_n}
=\infty$,} \vspace{2.00000pt}\label{bin-kon-1}  \\ e^{- \frac{e^{-\alpha ^*}}{(k-1)!}},
 &\hspace{-11.5pt}\text{ if $\lim_{n \to \infty}{\alpha_n}
=\alpha ^* \in (-\infty, \infty)$,}\vspace{2.00000pt} \label{bin-kon-e}
 \end{subnumcases}
  and
 \begin{align}
 &  \hspace{-39pt} \lim_{n \to \infty}\mathbb{P} \bigg[\hspace{-2pt}\begin{array}{l}\text{Graph }G_b(n,P_n,p_n)\text{
has a}\\\text{minimum node degree at least $k$}.\end{array}\hspace{-2pt}\bigg]  \nonumber
 \end{align}
 \vspace{2.00000pt}
 \begin{subnumcases}{ \quad =}
 0, &\hspace{-11.5pt}\text{ if $\lim_{n \to \infty}{\alpha_n}
=-\infty$}, \label{bin-kon-0-mnd} \vspace{2.00000pt}\\  1, &\hspace{-11.5pt}\text{ if $\lim_{n \to \infty}{\alpha_n}
=\infty$,} \label{bin-kon-1-mnd}\vspace{2.00000pt}  \\ e^{- \frac{e^{-\alpha ^*}}{(k-1)!}},
 &\hspace{-11.5pt}\text{ if $\lim_{n \to \infty}{\alpha_n}
=\alpha ^* \in (-\infty, \infty)$.}\vspace{2.00000pt} \label{bin-kon-e-mnd}
 \end{subnumcases}
 \end{thm}

 \begin{rem} \label{rm}

As we will explain in Section \ref{pfrig} within the proof of
Theorem \ref{thm:rig}, for (\ref{bin-kon-0}) (\ref{bin-kon-1}) (\ref{bin-kon-0-mnd}) (\ref{bin-kon-1-mnd}) (i.e., the zero--one laws), the condition $P_n =
\omega \big(n(\ln n)^5\big)$ can be weakened as $P_n = \Omega
\big(n(\ln n)^5\big)$, while we enforce $P_n = \omega \big(n(\ln
n)^5\big)$ for (\ref{bin-kon-e}) (\ref{bin-kon-e-mnd}).  

 \end{rem}

\subsubsection{$k$-Connectivity and Minimum Node Degree in Uniform Random Intersection
Graphs}~

For a uniform random intersection graph,
Theorem \ref{thm:urig}  below gives asymptotically exact probabilities for
$k$-connectivity and the property of minimum node degree being at least $k$. \vspace{-1.00000pt}

\begin{thm} \label{thm:urig} For a uniform random intersection graph
$G_u(n,P_n,K_n)$, with a sequence $\alpha_n$ for all $n $ defined through
\begin{align}
 \frac{{K_n}^2}{P_n} & = \frac{\ln  n + {(k-1)} \ln \ln n +
 {\alpha_n}}{n},  \label{thm:urig:pe}
\end{align}
 if $K_n = \Omega \big(\sqrt{\ln n}\hspace{2pt}\big)$,
  then
 \begin{align}
 &  \hspace{-25pt} \lim_{n \to \infty}\mathbb{P}
\big[\hspace{2pt}\text{Graph }G_u(n,P_n,K_n)\text{
is $k$-connected}.\hspace{2pt}\big] \nonumber
 \end{align}
 \vspace{-3.00000pt}
 \begin{subnumcases}{ \quad =}
 0, &\hspace{-11.5pt}\text{ if $\lim_{n \to \infty}{\alpha_n}
=-\infty$}, \label{uni-kon-0} \\  1, &\hspace{-11.5pt}\text{ if $\lim_{n \to \infty}{\alpha_n}
=\infty$,} \label{uni-kon-1}  \\ e^{- \frac{e^{-\alpha ^*}}{(k-1)!}},
 &\hspace{-11.5pt}\text{ if $\lim_{n \to \infty}{\alpha_n}
=\alpha ^* \in (-\infty, \infty)$,} \label{uni-kon-e}
 \end{subnumcases}
  and
 \begin{align}
 &  \hspace{-39pt} \lim_{n \to \infty}\mathbb{P} \bigg[\hspace{-2pt}\begin{array}{l}\text{Graph }G_u(n,P_n,K_n)\text{
has a}\\\text{minimum node degree at least $k$}.\end{array}\hspace{-2pt}\bigg]  \nonumber
 \end{align}
 \vspace{-1.00000pt}
 \begin{subnumcases}{ \quad =}
 0, &\hspace{-11.5pt}\text{ if $\lim_{n \to \infty}{\alpha_n}
=-\infty$}, \label{uni-kon-0-mnd} \\  1, &\hspace{-11.5pt}\text{ if $\lim_{n \to \infty}{\alpha_n}
=\infty$,} \label{uni-kon-1-mnd}  \\ e^{- \frac{e^{-\alpha ^*}}{(k-1)!}},
 &\hspace{-11.5pt}\text{ if $\lim_{n \to \infty}{\alpha_n}
=\alpha ^* \in (-\infty, \infty)$.} \label{uni-kon-e-mnd}
 \end{subnumcases}
\end{thm}

\subsubsection{$k$-Connectivity and Minimum Node Degree in General Random Intersection
Graphs}~

For a general random intersection graph,
Theorem \ref{thm:grig} below provides asymptotically exact probabilities for
$k$-connectivity and the property of minimum node degree being at least $k$.

\begin{thm} \label{thm:grig} Consider a general random intersection graph
$G(n,P_n,\mathcal {D}_n)$. Let $X_n$ be a random variable following
probability distribution $\mathcal {D}_n$. With a sequence $\alpha_n$
for all $n $ defined through
\begin{align}
 \frac{\big\{\mathbb{E}[X_n]\big\}^2}{P_n} & =
 \frac{\ln  n + {(k-1)} \ln \ln n + {\alpha_n}}{n},
 \label{thm:grig:pe}
\end{align}
 if $\mathbb{E}[X_n] = \Omega\big(\sqrt{\ln n}\hspace{2pt}\big)$ and $\text{Var}[X_n] =
o\mathlarger{\mathlarger{\big(}}\frac{\{\mathbb{E}[X_n]\}^2}{ n(\ln
n)^2 }\mathlarger{\mathlarger{\big)}}$,
 then
\begin{align}
 &  \hspace{-25pt} \lim_{n \to \infty}\mathbb{P}
\big[\hspace{2pt}\text{Graph }G(n,P_n,\mathcal {D}_n)\text{
is $k$-connected}.\hspace{2pt}\big] \nonumber
 \end{align}
 \vspace{-3.00000pt}
 \begin{subnumcases}{ \quad =}
 0, &\hspace{-11.5pt}\text{ if $\lim_{n \to \infty}{\alpha_n}
=-\infty$}, \label{grg-kon-0} \\  1, &\hspace{-11.5pt}\text{ if $\lim_{n \to \infty}{\alpha_n}
=\infty$,} \label{grg-kon-1}  \\ e^{- \frac{e^{-\alpha ^*}}{(k-1)!}},
 &\hspace{-11.5pt}\text{ if $\lim_{n \to \infty}{\alpha_n}
=\alpha ^* \in (-\infty, \infty)$,} \label{grg-kon-e}
 \end{subnumcases}
  and
 \begin{align}
 &  \hspace{-39pt} \lim_{n \to \infty}\mathbb{P} \bigg[\hspace{-2pt}\begin{array}{l}\text{Graph }G(n,P_n,\mathcal {D}_n)\text{
has a}\\\text{minimum node degree at least $k$}.\end{array}\hspace{-2pt}\bigg]  \nonumber
 \end{align}
 \vspace{-1.00000pt}
 \begin{subnumcases}{ \quad =}
 0, &\hspace{-11.5pt}\text{ if $\lim_{n \to \infty}{\alpha_n}
=-\infty$}, \label{grg-kon-0-mnd} \\  1, &\hspace{-11.5pt}\text{ if $\lim_{n \to \infty}{\alpha_n}
=\infty$,} \label{grg-kon-1-mnd}  \\ e^{- \frac{e^{-\alpha ^*}}{(k-1)!}},
 &\hspace{-11.5pt}\text{ if $\lim_{n \to \infty}{\alpha_n}
=\alpha ^* \in (-\infty, \infty)$.} \label{grg-kon-e-mnd}
 \end{subnumcases}

\end{thm}

\subsection{Asymptotic Zero--One Laws for  $k$-Robustness}
\label{sec:main:res:rb}

\subsubsection{$k$-Robustness in Binomial Random Intersection
Graphs}

Theorem \ref{thm:rig:rb} below gives an asymptotic zero--one law for
$k$-robustness in a binomial random intersection graph.

\begin{thm} \label{thm:rig:rb} For a binomial random intersection graph
$G_b(n,P_n,p_n)$, with a sequence $\alpha_n$ for all $n $ defined
through
\begin{align}
  {p_n}^2 P_n & =
 \frac{\ln  n + {(k-1)} \ln \ln n + {\alpha_n}}{n},  \label{thm:rig:pe:rb}
\end{align}
if $P_n = \Omega \big(n(\ln n)^5\big)$,
 then
\begin{align}
 & \lim_{n \to \infty}\mathbb{P} \big[\hspace{2pt}\text{Graph }G_b(n,P_n,p_n)\text{
is $k$-robust}.\hspace{2pt}\big] \nonumber
 \end{align}
 \vspace{-5pt}
 \begin{subnumcases}{ \quad =}
0, &\text{ if $\lim_{n \to \infty}{\alpha_n}
=-\infty$},\label{bin-krb-0} \\  1, &\text{ if $\lim_{n \to \infty}{\alpha_n}
=\infty$.}  \label{bin-krb-1}
 \end{subnumcases}

 \end{thm}

\subsubsection{$k$-Robustness in Uniform Random Intersection Graphs}

 Theorem \ref{thm:urig:rb} below presents an asymptotic zero--one law
for   $k$-robustness in a uniform random intersection
graph.

\begin{thm} \label{thm:urig:rb} For a uniform random intersection graph
$G_u(n,P_n,K_n)$, with a sequence $\alpha_n$ for all $n $ defined
through
\begin{align}
 \frac{{K_n}^2}{P_n} & = \frac{\ln  n + {(k-1)} \ln \ln n +
 {\alpha_n}}{n}, \label{thm:urig:pe:rb}
\end{align}
if $K_n = \Omega \big((\ln n)^3\big)$,
 then
\begin{align}
 & \lim_{n \to \infty}\mathbb{P} \big[\hspace{2pt}\text{Graph }G_u(n,P_n,K_n)\text{
is $k$-robust}.\hspace{2pt}\big] \nonumber
 \end{align}
\vspace{-5pt}
 \begin{subnumcases}{ \quad =}
0, &\text{ if $\lim_{n \to \infty}{\alpha_n}
=-\infty$},\label{uni-krb-0} \\  1, &\text{ if $\lim_{n \to \infty}{\alpha_n}
=\infty$.}  \label{uni-krb-1}
 \end{subnumcases}

\end{thm}

\subsubsection{$k$-Robustness in General Random Intersection
Graphs}

Theorem \ref{thm:grig:rb} as follows provides an asymptotic zero--one law for
 $k$-robustness in a general random intersection graph.

\begin{thm} \label{thm:grig:rb} Consider a general random intersection graph
$G(n,P_n,\mathcal {D}_n)$. Let $X_n$ be a random variable following
probability distribution $\mathcal {D}_n$. With a sequence $\alpha_n$
for all $n $ defined through
\begin{align}
 \frac{\big\{\mathbb{E}[X_n]\big\}^2}{P_n} & =
 \frac{\ln  n + {(k-1)} \ln \ln n + {\alpha_n}}{n},
 \label{thm:grig:pe:rb}
\end{align}
if $\mathbb{E}[X_n] = \Omega \big((\ln n)^3\big)$ and $\text{Var}[X_n] =
o\mathlarger{\mathlarger{\big(}}\frac{\{\mathbb{E}[X_n]\}^2}{ n(\ln
n)^2 }\mathlarger{\mathlarger{\big)}}$, then
\begin{align}
 & \lim_{n \to \infty}\mathbb{P} \big[\hspace{2pt}\text{Graph }G(n,P_n,\mathcal {D}_n)\text{
is $k$-robust}.\hspace{2pt}\big] \nonumber \\
& \quad =
\begin{cases} 0, &\text{ if $\lim_{n \to \infty}{\alpha_n}
=-\infty$}, \\  1, &\text{ if $\lim_{n \to \infty}{\alpha_n}
=\infty$.} \end{cases} \nonumber
 \end{align}

\end{thm}

In view of Theorems \ref{thm:rig}--\ref{thm:grig:rb}, for each
binomial/uniform/general random intersection graph, its
$k$-connectivity, $k$-robustness  and the property of minimum node degree being at least $k$ asymptotically obey the same
zero--one laws. Moreover, these zero--one laws are all \emph{sharp} since
$|\alpha_n|$ can be much smaller compared to $\ln n$; e.g., even
$\alpha_n = \pm \cdot \ln \ln \cdot
\cdot \cdot \ln n$ satisfies $\lim_{n \to
\infty}{\alpha_n} =\pm\infty$.

We compare our results of random intersection graphs with those of Erd\H{o}s--R\'{e}nyi graphs below. From \cite[Section 1.1]{PES:6114960}, ${p_n}^2 P_n$ in the scaling conditions (\ref{thm:rig:pe}) and (\ref{thm:rig:pe:rb}) of Theorems  \ref{thm:rig} and \ref{thm:rig:rb} is an asymptotics of the edge probability in a binomial random intersection graph
$G_b(n,P_n,p_n)$. Also, by \cite[Lemma 1]{bloznelis2013}, $\frac{{K_n}^2}{P_n}$ in the scaling conditions (\ref{thm:urig:pe}) and (\ref{thm:urig:pe:rb}) of Theorems  \ref{thm:urig} and  \ref{thm:urig:rb} (resp., $\frac{\{\mathbb{E}[X_n]\}^2}{P_n}$ in the scaling conditions (\ref{thm:grig:pe}) and (\ref{thm:grig:pe:rb}) of Theorems  \ref{thm:grig} and \ref{thm:grig:rb}  is an asymptotics of the edge probability in a uniform random intersection graph
$G_u(n,P_n,K_n)$ (resp., a general random intersection graph
$G(n,P_n,\mathcal {D}_n)$). Then comparing Theorems \ref{thm:rig}--\ref{thm:grig} with Lemma \ref{er_kconmnd}, and comparing Theorems \ref{thm:rig:rb}--\ref{thm:grig:rb} with Lemma \ref{er_robust}, we conclude   binomial/uniform/general random intersection graphs under certain parameter conditions\footnote{Under other parameter conditions, the conclusion may not hold
as in the case of binomial random intersection graphs shown by Rybarczyk \cite{zz,2013arXiv1301.0466R}.} exhibit the same behavior with Erd\H{o}s-R\'enyi graphs in the sense that for each of (i) $k$-connectivity, (ii) the property of minimum node degree being at least $k$, and (iii) $k$-robustness, a common point for the  transition from a zero-law to a one-law
occurs when the edge probability equals $\frac{\ln  n + {(k-1)} \ln \ln n}{n}$. The term $\alpha_n$ in Equations (1) (resp., (4) and (7)), or Equations (10) (resp., (12) and (14)) measures how much ${p_n}^2 P_n$ (resp., $\frac{{K_n}^2}{P_n}$ and $\frac{\{\mathbb{E}[X_n]\}^2}{P_n}$) is away from the critical value $\frac{\ln  n + {(k-1)} \ln \ln n}{n}$.

%

%
%


\subsection{A Discussion of Parameter Conditions} \label{ParameterConditions}

%

%



Note that we impose conditions on the parameters in the theorems; e.g., $P_n = \omega \big(n(\ln n)^5\big)$ in Theorem \ref{thm:rig}, and $K_n = \Omega \big(\sqrt{\ln n}\hspace{2pt}\big)$ in Theorem \ref{thm:urig}. These conditions are enforced to have the proofs get through and are not that conservative as explained below. We take a binomial random
intersection graph as an example and note that Theorem \ref{thm:rig} for $k$-connectivity in a binomial random
intersection graph does not hold if the condition $P_n = \omega \big(n(\ln n)^5\big)$ in Theorem \ref{thm:rig} is replaced by $P_n = n^{\tau}$ for a positive constant $\tau<1$. Specifically, we use \cite[Theorem 4]{zz} and \cite[Conjecture 1]{zz} confirmed by later work \cite{2013arXiv1301.0466R} to have the following claim:\vspace{-1pt}
  \begin{cla} \label{nPtaulem}
{Under $P_n = n^{\tau}$ for a positive constant $\tau<1$, with a sequence $\gamma_n$ for all $n $ defined
through\vspace{-1pt}
\begin{align}
p_n P_n =
 \ln  n  + {\gamma_n},\vspace{-1pt} \label{pnPnscaling}
 \end{align}
 then\vspace{-1pt}
\begin{align}
& \lim_{n \to \infty}\mathbb{P}
[G_b(n,P_n,p_n)\text{
is $k$-connected}.]
\nonumber \\ & \quad =\begin{cases} 0, &\text{ if\hspace{2pt}~$\lim_{n \to \infty}{\gamma_n}
=-\infty$}, \\  1, &\text{ if\hspace{2pt}~$\lim_{n \to \infty}{\gamma_n}
=\infty$.} \end{cases}\vspace{-1pt} \nonumber
\end{align}}
 \end{cla}
 Note that different from (\ref{thm:rig:pe}), the scaling condition (\ref{pnPnscaling}) above does not depend on $k$.

 Claim \ref{nPtaulem} has $P_n = n^{\tau}$ for a positive constant $\tau<1$, while Theorem \ref{thm:rig} has $P_n = \omega \big(n(\ln n)^5\big)$. We let $\delta$ denote an arbitrary constant with $\tau<\delta<\frac{\tau+1}{2}$ below. Claim \ref{nPtaulem} shows that the probability of $G_b(n,n^{\tau},n^{-\delta})$ (i.e., $G_b(n,P_n,p_n)$ with $P_n = n^{\tau}$ and $p_n = n^{-\delta}$) being $k$-connected asymptotically converges to $0$ since $\gamma_n$ specified by (\ref{pnPnscaling}) satisfies
 \begin{align}
\gamma_n & = p_n P_n - \ln  n = n^{\tau-\delta} - \ln  n \to -\infty, \text{ as $n \to \infty$}. \nonumber
 \end{align}
 In contrast, Theorem \ref{thm:rig} with $P_n = \omega \big(n(\ln n)^5\big)$ replaced by $P_n = n^{\tau}$ for a positive constant $\tau<1$ presents that the probability of $G_b(n,n^{\tau},n^{-\delta})$ (i.e., $G_b(n,P_n,p_n)$ with $P_n = n^{\tau}$ and $p_n = n^{-\delta}$) being $k$-connected asymptotically approaches to $1$
 because $\alpha_n$ defined by (\ref{thm:rig:pe}) satisfies
  \begin{align}
 \alpha_n & = n {p_n}^2 P_n - [\ln  n + {(k-1)} \ln \ln n] \nonumber \\ & = n^{1+\tau-2\delta} - [\ln  n + {(k-1)} \ln \ln n] \to \infty, \text{ as $n \to \infty$}. \nonumber
\end{align}
Hence, Claim \ref{nPtaulem} shows that if the condition $P_n = \omega \big(n(\ln n)^5\big)$ of Theorem \ref{thm:rig} is replaced by $P_n = n^{\tau}$ for a positive constant $\tau<1$, we will not obtain the $k$-connectivity result of Theorem \ref{thm:rig}. A future work is to investigate the intermediate range $\omega \big(n^{\tau}\big) = P_n = O\big(n(\ln n)^5\big)$.

We have discussed the parameter conditions for binomial random
intersection graphs. It is unclear whether $K_n = \Omega \big(\sqrt{\ln n}\hspace{2pt}\big)$ in Theorem \ref{thm:urig} and $K_n = \Omega \big((\ln n)^3\big)$ in Theorem \ref{thm:urig:rb} for uniform random
intersection graphs can be weakened since these conditions are also often enforced in related work \cite{Rybarczyk,ZhaoYaganGligor,yagan}. Moreover, these conditions are applicable to secure sensor networks since it has been shown that $K_n$ is at least on the order of $\ln n$ to have reasonable connectivity and resiliency \cite{YaganThesis,4198829,DiPietroTissec}. For a general random
intersection graph, Ya\u{g}an \cite{yagan2015zero} recently obtains a zero--one law for connectivity and  shows in \cite[Section 3.3]{yagan2015zero} that Theorem \ref{thm:grig} with $\text{Var}[X_n] =
o\mathlarger{\mathlarger{\big(}}\frac{\{\mathbb{E}[X_n]\}^2}{ n(\ln
n)^2 }\mathlarger{\mathlarger{\big)}}$ replaced by a broader condition does not hold.

To conclude, the parameter conditions in our theorems are not that conservative.

\section{Numerical Experiments} \label{sec:expe}

We present numerical experiments in the non-asymptotic regime to
confirm our theoretical results. 

\begin{figure}[!t]
  \centering
 \includegraphics[width=0.5\textwidth]{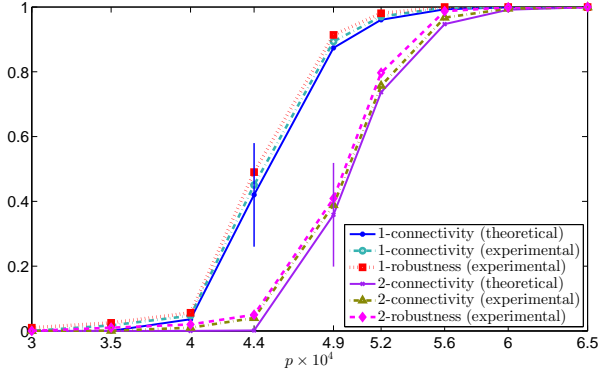}
\vspace{-17pt}\caption{A plot of the empirical probabilities that a binomial random
intersection graph $G_b(n,P,p)$ has $k$-connectivity or
$k$-robustness as a function of $p$, with $n=2000$, $P=20000$ and
$k=2,6$. \vspace{-9pt} } \label{fig}
\end{figure}

\begin{figure}[!t]
  \centering
 \includegraphics[width=0.5\textwidth]{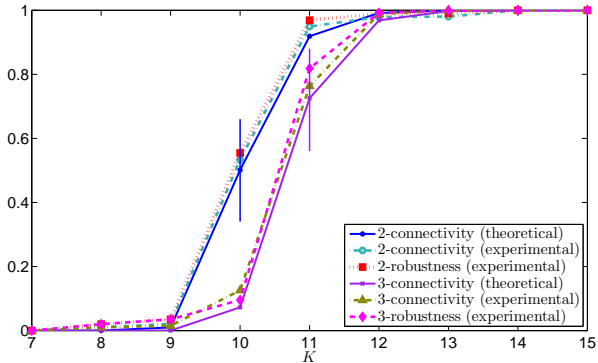}
\vspace{-17pt} \caption{A plot of the empirical probabilities that a uniform random intersection
graph $G_u(n,P,K)$ has $k$-connectivity or $k$-robustness as a
function of $K$, with $n=2000$, $P=20000$ and $k=3,4$. \vspace{13pt}
 }\label{figa}
\end{figure}

Figure \ref{fig} depicts the probability that a binomial random
intersection graph $G_b(n,P,p)$ has $k$-connectivity or
$k$-robustness, for $k = 1,2$. Similarly, Figure \ref{figa} illustrates the probability of $k$-connectivity or
$k$-robustness for
$k = 2, 3$ in a uniform random intersection graph $G_u(n,P,K)$. In
all set of experiments, we fix the number of nodes at $n=2000$ and
the object pool size $P = 20000$. For each pair $(n,P,p)$ (resp.,
$(n,P,K)$), we generate $1000$ independent samples of $G_b(n,P,p)$
(resp., $G_u(n,P,K)$) and count the number of times that the
obtained graphs are $k$-connected or $k$-robust. Then the counts
divided by $1000$ become the corresponding empirical probabilities.
As illustrated in Figures \ref{fig} and \ref{figa}, there is an
evident transition in the probabilities of $k$-connectivity
and $k$-robustness. Also, for each $k$, the curves of
$k$-connectivity and $k$-robustness are close to each other. Furthermore, the vertical lines in Figure \ref{fig} specify $p$ such that ${p}^2 P$ equals $\frac{\ln  n + {(k-1)} \ln \ln n}{n}$, while the vertical lines in Figure \ref{figa} specify $K$ such that $\frac{{K}^2}{P}$ is closest to $\frac{\ln  n + {(k-1)} \ln \ln n}{n}$ (since $K$ and $P$ are both integers, there might not exist $K$ satisfying $\frac{K^2}{P} = \frac{\ln  n + {(k-1)} \ln \ln n}{n}$).

The vertical lines in Figure \ref{fig} are at $4.4\times 10^{-4}$ and $4.9\times 10^{-4}$ because under \vspace{.5pt} $n=2000$ and $P=20000$, $\sqrt{\frac{\ln  n + {(k-1)} \ln \ln n}{nP}}$ is $\sqrt{\frac{\ln 2000}{2000 \times 20000}}\approx 4.4\times 10^{-4}$ \vspace{1pt} for $k=1$ and is $\sqrt{\frac{\ln 2000 + \ln \ln 2000}{2000 \times 20000}}\approx 4.9\times 10^{-4}$ for $k=2$. The vertical lines in Figure \ref{figa} are at $10$ and $11$ because \vspace{.5pt} under $n=2000$ and $P=20000$, \vspace{2pt} $\argmin_{K} \big| \frac{K^2}{P} - \frac{\ln  n + {(k-1)} \ln \ln n}{n}\big|$ equals $10$ for $k=2$ from $\frac{\ln 2000 + \ln \ln 2000}{2000}\approx 0.00481$, $\frac{9^2}{20000}\approx 0.00405$ and $\frac{10^2}{20000}\approx 0.005$, and equals $11$ for $k=3$ from $\frac{\ln 2000 + 2\ln \ln 2000}{2000}\approx 0.00583$, $\frac{10^2}{20000}\approx 0.005$ and  $\frac{11^2}{20000}\approx 0.00605$.

\section{Auxiliary Lemmas} \label{sec:factlem}

We present lemmas that are used in proving the
theorems.

 \subsection{Relationships between $k$-Robustness, $k$-Connectivity, and Minimum Node Degree}

 Lemma \ref{lem-k-robu-mnd} below, taken from \cite[Lemma 1]{6425841}, provides relationships between $k$-robustness, $k$-connectivity, and minimum node degree.

\begin{lem}[\hspace{-.2pt}{\cite[Lemma 1]{6425841}}\hspace{0pt}] \label{lem-k-robu-mnd}
 For any graph/network, $k$-robustness implies $k$-connectivity, which further implies that the minimum node degree is at least $k$.
\end{lem}

To prove that $k$-robustness implies $k$-connectivity,
\cite[Lemma 1]{6425841} shows that a graph $G$ that is not $k$-connected is also not $k$-robust. The idea is that for $G$ being not $k$-connected, there exists a set of $k-1$ nodes, whose deletion gives two disjoint subgraphs with node sets $V_a$ and $V_b$, respectively. Then in graph $G$, each node in  $V_a$ has less than $k$ neighbors outside of  $V_a$, and each node in  $V_b$ has less than $k$ neighbors outside of  $V_b$, so $G$ is not $k$-robust. Then it follows that $k$-robustness implies $k$-connectivity. In addition, it is clear that $k$-connectivity implies that the minimum node degree is at least $k$.

 Lemma \ref{lem-k-robu-mnd} is used in the proofs of Theorems \ref{thm:rig:rb} and \ref{thm:urig:rb}.

\subsection{Results of Erd\H{o}s-R\'enyi Graphs}

Lemma \ref{er_kconmnd} below  by  Erd\H{o}s and R\'enyi \cite{erdos61conn} investigates $k$-connectivity and minimum node degree in Erd\H{o}s-R\'enyi graphs.  An Erd\H{o}s--R\'{e}nyi graph
$G(n,\hat{p}_n)$ \cite{citeulike:4012374} is defined on a set of $n$
nodes such that any two nodes have an edge in between independently
with probability $\hat{p}_n$.

\begin{lem}[{Erd\H{o}s and R\'enyi} \cite{erdos61conn}]
\label{er_kconmnd} For an Erd\H{o}s--R\'{e}nyi graph
$G(n,\hat{p}_n)$, with a sequence $\alpha_n$ for all $n$ through
\begin{align}
\hat{p}_n = & \frac{\ln  n + {(k-1)} \ln \ln n + {\alpha_n}}{n} , \nonumber
 \end{align}
then it holds that
\begin{align}
 & \lim_{n \to \infty} \hspace{-1pt}\mathbb{P} \big[G(n,\hat{p}_n)\text{
is $k$-connected}.\big] \nonumber \\  & =
 \lim_{n \to \infty} \hspace{-1pt}\mathbb{P} \big[G(n,\hat{p}_n)\text{
has a minimum node degree at least $k$}.\big]\nonumber \\ &= \begin{cases} 0, \text{ if $\lim_{n \to
\infty}{\alpha_n} \hspace{-2pt}=\hspace{-2pt}-\infty$}, \\  1,
\text{ if $\lim_{n \to \infty}{\alpha_n}
\hspace{-2pt}=\hspace{-2pt}\infty$.}
\end{cases}  \nonumber
 \end{align}
\end{lem}

Lemma \ref{er_kconmnd} is used in the comparison of random intersection graphs and Erd\H{o}s--R\'{e}nyi graphs.

Lemma \ref{er_robust} below presents the result on $k$-robustness in Erd\H{o}s-R\'enyi graphs.

\begin{lem}
\label{er_robust} For an Erd\H{o}s--R\'{e}nyi graph
$G(n,\hat{p}_n)$, with a sequence $\alpha_n$ for all $n$ through
\begin{align}
\hat{p}_n = & \frac{\ln  n + {(k-1)} \ln \ln n + {\alpha_n}}{n},
 \nonumber
 \end{align}
then it holds that
\begin{align}
 & \lim_{n \to \infty} \hspace{-1pt}\mathbb{P} \big[G(n,\hat{p}_n)\text{
is $k$-robust}.\big]  = \begin{cases} 0, \text{ if $\lim_{n \to
\infty}{\alpha_n} \hspace{-2pt}=\hspace{-2pt}-\infty$}, \\  1,
\text{ if $\lim_{n \to \infty}{\alpha_n}
\hspace{-2pt}=\hspace{-2pt}\infty$.}
\end{cases}  \label{er_rb}
 \end{align}
\end{lem} 

Lemma \ref{er_robust} is applied to Section \ref{prf:thm:rig:rb} for proving Theorem \ref{thm:rig:rb}. Lemma \ref{er_robust} is also used in the comparison of random intersection graphs and Erd\H{o}s--R\'{e}nyi graphs.

To prove Lemma \ref{er_robust}, we note the following three
facts. (a) The desired result
(\ref{er_rb}) with $|\alpha_n| = o(\ln \ln n)$ is demonstrated in
\cite[Theorem 3]{6425841}. (b) By \cite[Facts 3 and 7]{zz}, for any
monotone increasing graph property $\mathcal {I}$, the probability
that graph $G(n,\hat{p}_n)$ has property $\mathcal {I}$ is
non-decreasing as $\hat{p}_n$ increases, where a graph
property is called monotone increasing if it holds under the
addition of edges. (c) $k$-Robustness is a monotone increasing graph property
according to \cite[Lemma 3]{6481629}. In view of (a) (b) and (c)
above, we obtain Lemma \ref{er_robust}. 

\subsection{Lemmas for Graph Coupling}

We present several lemmas for graph coupling below. Formally, a coupling
\cite{zz,2013arXiv1301.0466R,ZhaoCDC} of two random graphs
$G_1$ and $G_2$ means a probability space on which random graphs
$G_1'$ and $G_2'$ are defined such that $G_1'$ and $G_2'$ have the
same distributions as $G_1$ and $G_2$, respectively. If $G_1'$ is a spanning
subgraph (resp., spanning supergraph) of $G_2'$, we say that under the graph coupling, $G_1$ is a spanning subgraph (resp., spanning supergraph) of $G_2$, where a spanning subgraph (resp., spanning supergraph) is a subgraph (resp., supergraph) that has the same node set with the original graph.



 Following Rybarczyk's notation \cite{zz}, we
write
\begin{align}
G_1 \preceq  G_2 \quad (\text{resp.}, G_1 \preceq_{1-o(1)} G_2)
\label{g1g2coupling}
\end{align}
if there exists a coupling under which $G_1$ is a {spanning subgraph}
of $G_2$ with probability $1$ (resp., $1-o(1)$).
 We
write
\begin{align}
G_2 \succeq  G_1 \quad (\text{resp.}, G_2 \succeq_{1-o(1)} G_1)
\label{g2g1coupling}
\end{align}
if there exists a coupling under which $G_2$ is a {spanning supergraph}
of $G_1$ with probability $1$ (resp., $1-o(1)$).
 According to the definitions above, $G_1 \preceq  G_2$ and $G_2 \succeq  G_1$ are equivalent, while $G_1 \preceq_{1-o(1)} G_2$ and $G_2 \succeq_{1-o(1)} G_1$ are equivalent.

In view that $k$-connectivity and $k$-robustness
are monotone increasing graph properties \cite{erdos61conn,6481629}, it is natural to obtain that under $G_1 \preceq  G_2$ or $G_1 \preceq_{1-o(1)} G_2$, if $G_1$ is $k$-connected (resp. $k$-robust) with a probability at least $1-o(1)$, then $G_2$ is also $k$-connected (resp. $k$-robust) with a probability at least $1-o(1)$. This result is formally presented in
Lemma \ref{mono-gcp} below given by Rybarczyk \cite{zz}. Lemma \ref{mono-gcp} considers any monotone increasing graph property for generality.

\begin{lem}[Rybarczyk \cite{zz}]  \label{mono-gcp}

For two random graphs $G_1$ and $G_2$, the following results hold
for any monotone increasing graph property $\mathcal {I}$.

\begin{itemize}
  \item[\textbf{(a)}] If $G_1 \preceq G_2$, then $$\mathbb{P} \big[ G_2\text{
has $\mathcal {I}$.}\big] \geq \mathbb{P} \big[ G_1\text{ has
$\mathcal {I}$.}\big].$$
  \item[\textbf{(b)}] If $G_1 \preceq_{1-o(1)}  G_2$, then $$\mathbb{P} \big[ G_2\text{
has $\mathcal {I}$.}\big] \geq \mathbb{P} \big[ G_1\text{ has
$\mathcal {I}$.}\big]- o(1).$$
\end{itemize}

\end{lem}

Lemma \ref{mono-gcp} is used in many places of this paper. We then present Lemmas \ref{lem:cp}--\ref{cp_urig_rig}. Except Lemma \ref{rkgikg} which is from \cite[Lemma 4]{Rybarczyk}, the proofs of other lemmas are deferred to
 Section \ref{sec:prf:fact:lem}.

\subsubsection{Coupling between general random intersection graphs and uniform random intersection graphs}  \label{secffalem:cp}

\begin{lem} \label{lem:cp}
Let $X_n$ be a random variable with \vspace{2pt}probability
distribution $\mathcal {D}_n$. If $\text{Var}[X_n] =
o\mathlarger{\mathlarger{\big(}}\frac{\{\mathbb{E}[X_n]\}^2}{n (\ln
n)^2 }\mathlarger{\mathlarger{\big)}}$, then there exists
$\epsilon_n = o\big(\frac{1}{\ln  n}\big)$ such that
\begin{align}
G_u\big(n, P_n,
 (1  -& \epsilon_n) \mathbb{E}[X_n]\big)   \preceq_{1-o(1)} G(n,P_n,\mathcal {D}_n)  \nonumber \\
 & \preceq_{1-o(1)} G_u\big(n, P_n,
 (1 + \epsilon_n)\mathbb{E}[X_n]\big). \label{cptac15new}
\end{align}

\end{lem} 


 Lemma \ref{lem:cp} is shown in Section \ref{Couplinggeneraluniform}, and is used to prove Theorems \ref{thm:grig} and \ref{thm:grig:rb}.



\subsubsection{Coupling between binomial random intersection graphs and Erd\H{o}s--R\'{e}nyi graphs} \label{secfagcp_rig_er}

\begin{lem} \label{cp_rig_er}

If $p_n = O\left( \frac{1}{n\ln n} \right)$\vspace{3pt} and ${p_n}^2
P_n =
 O\left( \frac{1}{\ln n} \right)$, then there exists $\hat{p}_n =
{p_n}^2 P_n \cdot \left[1-
 O\left(\frac{1}{ \ln n}\right)\right]$ such that  
 \begin{align}
G(n,\hat{p}_n) \preceq_{1-o(1)} G_b(n,P_n,p_n). \label{cp_res_rig_er}
\end{align}
\end{lem} 

 Lemma \ref{cp_rig_er} is shown in Section \ref{CouplingbinomialER}, and is used to prove Theorem \ref{thm:rig:rb}.


\subsubsection{Coupling between binomial random intersection graphs and uniform random intersection graphs} \label{secacp_rkgikg}

\begin{lem}[{\hspace{-0.2pt}{\cite[Lemma 4]{Rybarczyk}}}] \label{rkgikg} If $p_n P_n = \omega\left( \ln n \right)$, and for all $n$
sufficiently large,
\begin{align}
K_{n,-}  & \leq p_n P_n - \sqrt{3(p_n P_n + \ln n) \ln n} ,
\nonumber \\
K_{n,+}  & \geq p_n P_n + \sqrt{3(p_n P_n + \ln n) \ln n}  ,
\nonumber
\end{align}
then
\begin{align}
G_u(n,P_n,K_{n,-})
& \preceq_{1-o(1)}
 G_b(n,P_n,p_n) \nonumber \\
& \preceq_{1-o(1)}
G_u(n,P_n,K_{n,+}).   \nonumber
\end{align}
\end{lem}  

 Lemma \ref{rkgikg} is used in the proof of Theorem \ref{thm:rig}.

\begin{lem} \label{cp_urig_rig}

If $K_n \hspace{-1pt}=\hspace{-1pt} \omega\left( \ln n \right)$ and
$p_n \hspace{-1pt}=\hspace{-1pt} \frac{K_n}{P_n}
 \left(1 - \sqrt{\frac{3\ln
n}{K_n }}\hspace{2pt}\right)$, then   
\begin{align}
G_u(n,P_n,K_n) \succeq_{1-o(1)} G_b(n,P_n,p_n). \nonumber
\end{align}
\end{lem} 


Lemma \ref{cp_urig_rig} is established in Section \ref{Couplingcp_urig_rig}, and is used to prove Theorem \ref{thm:urig:rb}.

%



We will use each of Lemmas \ref{lem:cp}--\ref{cp_urig_rig} along with Lemma \ref{mono-gcp}. For simplicity, we just use Lemma \ref{lem:cp} as an example to explain its implication with Lemma \ref{mono-gcp}. From property (b) of Lemma \ref{mono-gcp} and result (\ref{cptac15new}) of Lemma \ref{lem:cp}, we obtain for any monotone increasing graph property $\mathcal {I}$ that
\begin{align}
 & \mathbb{P} \big[\hspace{2pt}\text{Graph }G_u(n, P_n,
 (1 - \epsilon_n)\mathbb{E}[X_n])\text{ has $\mathcal {I}$}.
  \hspace{2pt}\big]  - o(1)
 \nonumber \\
&  \leq \mathbb{P} \big[\hspace{2pt}\text{Graph }G(n,P_n,\mathcal {D}_n)\text{
has $\mathcal {I}$}.\hspace{2pt}\big] \nonumber \\
&  \leq \mathbb{P} \big[\hspace{2pt}\text{Graph }G_u(n, P_n,
 (1 + \epsilon_n)\mathbb{E}[X_n])\text{ has $\mathcal {I}$}.
  \hspace{2pt}\big]  + o(1).\nonumber
 \end{align}

\section{Establishing Theorems \ref{thm:rig} and \ref{thm:grig}}
\label{sec:thmprf:kcon}

Theorems \ref{thm:rig}--\ref{thm:grig} describe results on
$k$-connectivity for binomial/uniform/general random intersection graphs. We prove
Theorems \ref{thm:rig} and \ref{thm:grig} in this section, and present the proof of Theorem \ref{thm:urig} separately as Section \ref{prf:thm:urig} next due to the length of the proof.

We briefly explain the idea of proving Theorems \ref{thm:rig} and \ref{thm:grig} from Theorem \ref{thm:urig} below. First, we demonstrate Theorem \ref{thm:rig} from Theorem \ref{thm:urig} using the coupling between binomial random intersection graphs and uniform random intersection graphs given by Lemma \ref{rkgikg} of Section \ref{secacp_rkgikg}. Second, we establish Theorem \ref{thm:grig} from Theorem \ref{thm:urig} using the coupling between general random intersection graphs and uniform random intersection graphs given by Lemma \ref{lem:cp} of Section \ref{secffalem:cp}.




\subsection{The Proof of Theorem \ref{thm:rig}} \label{pfrig}

As explained in Appendix \ref{seca:conf:bin}, we can introduce an extra condition $|\alpha_n| = o(\ln n)$ in proving Theorem \ref{thm:rig}. Then
from  Theorem \ref{thm:urig},  Lemmas \ref{mono-gcp} and \ref{rkgikg}, and
and the fact that both $k$-connectivity and the property of minimum node degree being at least $k$ are  monotone increasing graph properties, the proof of
Theorem \ref{thm:rig} is completed once we show that with
$K_{n,\pm}$ given by
\begin{align}
K_{n,\pm}  & = p_n P_n \pm \sqrt{3(p_n P_n + \ln n) \ln n} ,
\label{Kngeqsb}
\end{align}
under conditions of Theorem \ref{thm:rig} and $|\alpha_n| = o(\ln n)$, we have $K_{n,\pm}    =
\Omega \big(\sqrt{\ln n}\hspace{2pt}\big)$ and with $\alpha_{n,\pm}$
defined by
\begin{align}
 \frac{{K_{n,\pm}}^2}{P_n} & = \frac{\ln  n + {(k-1)} \ln \ln n +
 {\alpha_{n,\pm}}}{n},  \label{thm:urig:peab}
 \end{align}
then
\begin{align}
\alpha_{n,\pm}  &   = \alpha_{n} \pm o(1).  \label{thm:urig:peaaph}
\end{align}

From conditions (\ref{thm:rig:pe}) and $|\alpha_n| = o(\ln n)$,
 it is clear that
\begin{align}
 {p_n}^2 P_n & \sim
 \frac{\ln  n}{n} .\label{thm:rig:pe_sim}
\end{align}
Substituting (\ref{thm:rig:pe_sim}) and condition $P_n = \omega
\big(n(\ln n)^5\big)$ into (\ref{Kngeqsb}), we obtain
\begin{gather}
p_n P_n  = \sqrt{{p_n}^2 P_n \cdot P_n} = \omega
\bigg( \frac{\ln  n}{n} \cdot n(\ln n)^5\bigg) = \omega \big((\ln n)^3\big) , \nonumber
\\
K_{n,\pm}    = \omega \big((\ln n)^3\big) = \Omega \big(\sqrt{\ln
n}\hspace{2pt}\big) , \label{zh1}
\end{gather}
and
\begin{align}
 \frac{{K_{n,\pm}}^2}{P_n} & =  {p_n}^2 P_n \cdot \Bigg[1  \pm \sqrt{3\bigg(1 + \frac{\ln n}{p_n P_n}\bigg) \frac{\ln n}{p_n P_n}}\,\Bigg]  \nonumber \\ & =    {p_n}^2 P_n \cdot \bigg[1 \pm o\bigg(\frac{1}{\ln n}\bigg)\bigg]. \label{thm:urig:pea}
\end{align}
Then from (\ref{thm:rig:pe}) (\ref{thm:urig:peab}) and
(\ref{thm:urig:pea}),
  we obtain
(\ref{thm:urig:peaaph}). As explained before, with
(\ref{thm:urig:peab}) (\ref{thm:urig:peaaph}) and (\ref{zh1}),
Theorem \ref{thm:rig} is proved from Theorem
\ref{thm:urig} and Lemmas \ref{mono-gcp} and \ref{rkgikg}.

Finally, as noted in Remark \ref{rm} after  Theorem \ref{thm:rig}, to prove
 zero--one laws (\ref{bin-kon-0}) (\ref{bin-kon-1}) (\ref{bin-kon-0-mnd}) (\ref{bin-kon-1-mnd}) but not (\ref{bin-kon-e}) (\ref{bin-kon-e-mnd}) in Theorem \ref{thm:rig},  condition $P_n = \omega
\big(n(\ln n)^5\big)$ can be weakened as $P_n = \Omega \big(n(\ln
n)^5\big)$. This is seen by the argument that under $P_n =
\Omega \big(n(\ln n)^5\big)$, $K_{n,\pm}    =
\Omega \big(\sqrt{\ln n}\hspace{2pt}\big)$ still holds and (\ref{thm:urig:peaaph}) is
weakened as $\alpha_{n,\pm}   = \alpha_{n} \pm O(1)$, so we still have zero--one laws (\ref{bin-kon-0}) (\ref{bin-kon-1}) (\ref{bin-kon-0-mnd}) (\ref{bin-kon-1-mnd}).

\subsection{The Proof of Theorem \ref{thm:grig}}

Given Lemmas \ref{mono-gcp} and \ref{lem:cp} and the fact that both $k$-connectivity and the property of minimum node degree being at least $k$ are  monotone increasing graph properties,
 we will show Theorem \ref{thm:grig} 
 once proving   for any $\epsilon_n = o\left(\frac{1}{\ln  n}\right)$ that
\begin{align}
 & \lim_{n \to \infty}\mathbb{P} \big[ G_u(n, P_n,
 (1 \pm \epsilon_n)\mathbb{E}[X_n])
 \text{
is $k$-connected}. \big] \nonumber  \\
&   =
\begin{cases} 0, &\text{ if $\lim_{n \to \infty}{\alpha_n}
=-\infty$}, \\  1, &\text{ if $\lim_{n \to \infty}{\alpha_n}
=\infty$,} \\ e^{- \frac{e^{-\alpha ^*}}{(k-1)!}},
 &\text{ if $\lim_{n \to \infty}{\alpha_n}
=\alpha ^* \in (-\infty, \infty)$,} \end{cases} \label{kconn}
 \end{align}
 and
 \begin{align}
 & \lim_{n \to \infty}\mathbb{P} \bigg[\hspace{-2pt}\begin{array}{l}\text{Graph }G_u(n, P_n,
 (1 \pm \epsilon_n)\mathbb{E}[X_n])\text{
has a}\\\text{minimum node degree at least $k$}.\end{array}\hspace{-2pt}\bigg] \nonumber  \\
&   =
\begin{cases} 0, &\text{ if $\lim_{n \to \infty}{\alpha_n}
=-\infty$}, \\  1, &\text{ if $\lim_{n \to \infty}{\alpha_n}
=\infty$,} \\ e^{- \frac{e^{-\alpha ^*}}{(k-1)!}},
 &\text{ if $\lim_{n \to \infty}{\alpha_n}
=\alpha ^* \in (-\infty, \infty)$.} \end{cases} \label{kconn-mdn}
 \end{align}

 Under $\mathbb{E}[X_n] = \Omega \big(\sqrt{\ln n}\hspace{2pt}\big)$ and
$\epsilon_n = o\left(\frac{1}{\ln n}\right)$, it \vspace{1.5pt} follows that $(1 \pm \epsilon_n)\mathbb{E}[X_n]
 =  \Omega \big(\sqrt{\ln n}\hspace{2pt}\big)$. From Theorem \ref{thm:urig}, we will have (\ref{kconn}) and  (\ref{kconn-mdn}) once we prove that sequences $\gamma_n^{+}$ and $\gamma_n^{-}$ defined through
 \begin{align}
 \frac{\big\{(1\pm \epsilon_n)\mathbb{E}[X_n]\big\}^2}{P_n}
   & = \frac{\ln  n + {(k-1)}
 \ln \ln n + {\gamma_n^{\pm}}  }{n}\label{pe_epsilon4tac}
 \end{align}
 satisfy
  \begin{align}
\lim_{n \to \infty} \gamma_n^{\pm} &   =
\begin{cases} -\infty, &\text{ if $\lim_{n \to \infty}{\alpha_n}
=-\infty$}, \\  \infty, &\text{ if $\lim_{n \to \infty}{\alpha_n}
=\infty$,} \\ \alpha ^* ,
 &\text{ if $\lim_{n \to \infty}{\alpha_n}
=\alpha ^* \in (-\infty, \infty)$.} \end{cases} \label{pe_epsilon4tac2}
 \end{align}
 Now we establish (\ref{pe_epsilon4tac2}). From (\ref{thm:grig:pe})   (\ref{pe_epsilon4tac}) and $\epsilon_n = o\left(\frac{1}{\ln  n}\right)$, it follows that
 \begin{align}
\gamma_n^{\pm} & =   n \cdot  \frac{\big\{(1\pm \epsilon_n)\mathbb{E}[X_n]\big\}^2}{P_n} - [\ln  n + {(k-1)}
 \ln \ln n ] \nonumber \\
 & = (1\pm \epsilon_n)^2  [\ln  n + {(k-1)}
 \ln \ln n +\alpha_n ]  \nonumber \\
 &  \quad -  [\ln  n + {(k-1)}
 \ln \ln n ] \nonumber \\
 & = \alpha_n + \epsilon_n({\epsilon_n}\pm 2)   [\ln  n + {(k-1)}
 \ln \ln n +\alpha_n ] \nonumber \\
 & = \alpha_n \pm \bigg[o\bigg(\frac{\alpha_n}{\ln n}\bigg)+o(1)\bigg], \label{pe_epsilon4tac2tc}
 \end{align}
 where the last step uses $\epsilon_n = o\left(\frac{1}{\ln  n}\right)$.
 Then (\ref{pe_epsilon4tac2tc}) clearly implies (\ref{pe_epsilon4tac2}). Therefore, as mentioned above,
we establish (\ref{kconn}) (\ref{kconn-mdn}) and finally  Theorem \ref{thm:grig}.

\section{The Proof of Theorem \ref{thm:urig}}
\label{prf:thm:urig}

 As explained in Appendix \ref{seca:conf:unig}, we can introduce an extra condition $|\alpha_n| = o(\ln n)$ in proving Theorem \ref{thm:urig}. Then
since a necessary condition for a graph to be $k$-connected is that the minimum degree is at least $k$,  (\ref{uni-kon-0-mnd}) implies  (\ref{uni-kon-0}), and we have
 \begin{align}
&\mathbb{P}
\big[\hspace{2pt}\text{Graph }G_u(n,P_n,K_n)\text{
is $k$-connected}.\hspace{2pt}\big] \nonumber
\\ & =  \mathbb{P} \bigg[\hspace{-2pt}\begin{array}{l}\text{Graph }G_u(n,P_n,K_n)\text{
has a}\\\text{minimum node degree at least $k$}.\end{array}\hspace{-2pt}\bigg]   \nonumber
\\ & \quad -   \mathbb{P}\bigg[\hspace{-2pt}\begin{array}{l}G_u(n, P_n, K_n)\text{ has a minimum degree}\\\text{at least
}k,\text{ but is not $k$-connected}.\end{array}\hspace{-2pt}\bigg]. \label{kconvsmnd}
 \end{align}
 From (\ref{kconvsmnd}),  we know that (\ref{uni-kon-1}) (resp., (\ref{uni-kon-e})) will follow from the combination of Lemma  \ref{lemma-1} below and (\ref{uni-kon-1-mnd}) (resp., (\ref{uni-kon-e-mnd})), where Lemma  \ref{lemma-1} uses the extra condition $|\alpha_n| = o(\ln n)$ explained above. Also as mentioned before,
(\ref{uni-kon-0-mnd}) implies  (\ref{uni-kon-0}). Therefore, the proof of Theorem \ref{thm:urig} will be completed once we demonstrate (\ref{uni-kon-0-mnd}) (\ref{uni-kon-1-mnd})  (\ref{uni-kon-e-mnd}) and Lemma  \ref{lemma-1}, where we also use the extra condition $|\alpha_n| = o(\ln n)$ in proving   (\ref{uni-kon-0-mnd}) (\ref{uni-kon-1-mnd})  (\ref{uni-kon-e-mnd}). We let $e^{-\infty} = 0$ and $e^{\infty} = \infty$, so $e^{- \frac{e^{-\lim_{n \to \infty}{\alpha_n}}}{(k-1)!}}$  equals $0$ if $\lim_{n \to \infty}{\alpha_n}
=-\infty$, $1$ if $\lim_{n \to \infty}{\alpha_n}
=\infty$ and $e^{- \frac{e^{-\alpha ^*}}{(k-1)!}}$ if $\lim_{n \to \infty}{\alpha_n}
=\alpha ^* \in (-\infty, \infty)$. Then  (\ref{uni-kon-0-mnd}) (\ref{uni-kon-1-mnd})  (\ref{uni-kon-e-mnd}) under $|\alpha_n| = o(\ln n)$ can be compactly presented by Lemma  \ref{lemma-2} below. Hence, the proof of Theorem \ref{thm:urig} finally reduces to proving Lemmas  \ref{lemma-1} and  \ref{lemma-2}.


%
%

%

\begin{lem} 
 \label{lemma-1}
For a uniform random intersection graph $G_u(n, P_n, K_n)$ under $K_n = \Omega \big(\sqrt{\ln n}\hspace{2pt}\big)$ and $ \frac{{K_n}^2}{P_n}  = \frac{\ln  n + {(k-1)} \ln \ln n +
 {\alpha_n}}{n}$, where $\lim_{n \to \infty} \alpha_n$ exists and $|\alpha_n| = o(\ln n)$, it follows that
\begin{align}
 \lim_{n \to \infty} \hspace{-2pt} \mathbb{P}\bigg[\hspace{-4pt}\begin{array}{l}G_u(n, P_n, K_n)\text{ has a minimum degree}\\\text{at least
}k,\text{ but is not $k$-connected}.\end{array}\hspace{-4pt}\bigg] & \hspace{-1pt}=\hspace{-1pt} 0. \label{tbprof}
  \end{align}
\end{lem}


\begin{lem}  \label{lemma-2}
For a uniform random intersection graph $G_u(n, P_n, K_n)$ under $K_n = \Omega \big(\sqrt{\ln n}\hspace{2pt}\big)$ and $ \frac{{K_n}^2}{P_n}  = \frac{\ln  n + {(k-1)} \ln \ln n +
 {\alpha_n}}{n}$, where $\lim_{n \to \infty} \alpha_n$ exists and $|\alpha_n| = o(\ln n)$, it follows that
\begin{align}
&  \lim_{n \to \infty}  \mathbb{P}
\left[\hspace{2pt}G_u(n, P_n, K_n)\text{ has a minimum degree at least
}k.\hspace{2pt}\right]   \nonumber \\
&   \quad  = e^{- \frac{e^{-\lim_{n \to \infty}{\alpha_n}}}{(k-1)!}}. \nonumber 
 \end{align}
\end{lem}

To prove Lemma  \ref{lemma-1}, we use the following Lemma  \ref{lemma-1rp} on $G_u(n, P_n, K_n) \bcap G(n,\hat{p}_n)$, where $G(n,\hat{p}_n)$ is an
  Erd\H{o}s--R\'{e}nyi graph with $n$ nodes and edge probability $\hat{p}_n$, and the intersection of two graphs $G_A$ and $G_B$ defined on the same node set is constructed on the node set with the edge set being the intersection of the edge sets of $G_A$ and $G_B$.

\begin{lem} [\hspace{0pt}{Our work \cite[Propositions 3 and 4]{ZhaoYaganGligor}}\hspace{0pt}]
 \label{lemma-1rp}
For a graph $G_u(n, P_n, K_n) \bcap G(n,\hat{p}_n)$ under $P_n = \Omega (n)$, $\frac{K_n}{P_n} = o(1)$ and $ \frac{{K_n}^2}{P_n} \cdot \hat{p}_n  = \frac{\ln  n + {(k-1)} \ln \ln n +
 {\alpha_n}}{n}$, where $\lim_{n \to \infty} \alpha_n$ exists and $|\alpha_n| = o(\ln n)$, it follows that
\begin{align}
 \lim_{n \to \infty} \hspace{-2pt} \mathbb{P} \bigg[\hspace{-4pt}\begin{array}{l}G_u(n, P_n, K_n)\bcap G(n,\hat{p}_n)\text{ has a minimum}\\\text{degree at least }k,\text{ but is not $k$-connected}.\end{array}\hspace{-4pt}\bigg] & \hspace{-1pt}=\hspace{-1pt} 0.
  \end{align}
\end{lem}

Lemma \ref{lemma-1rp} is from our work \cite[Propositions 3 and 4]{ZhaoYaganGligor}. Setting $\hat{p}_n = 1$, we have  $G_u(n, P_n, K_n) \bcap G(n,\hat{p}_n) = G_u(n, P_n, K_n)$ and obtain results on $G_u(n, P_n, K_n)$ from Lemma  \ref{lemma-1rp}:
\begin{itemize}
\item[] \textit{For $G_u(n, P_n, K_n)$ under $P_n = \Omega (n)$, $\frac{K_n}{P_n} = o(1)$ and $ \frac{{K_n}^2}{P_n}   = \frac{\ln  n + {(k-1)} \ln \ln n +
 {\alpha_n}}{n}$, where $\lim_{n \to \infty} \alpha_n$ exists and $|\alpha_n| = o(\ln n)$, result (\ref{tbprof}) holds.}
\end{itemize}
Then clearly, Lemma  \ref{lemma-1} will be proved once we show conditions in Lemma  \ref{lemma-1} imply $P_n = \Omega (n)$ and $\frac{K_n}{P_n} = o(1)$. From conditions in Lemma  \ref{lemma-1}, we have $K_n = \Omega \big(\sqrt{\ln n}\hspace{2pt}\big)$ and $ \frac{{K_n}^2}{P_n}  = \frac{\ln  n + {(k-1)} \ln \ln n +
 {\alpha_n}}{n} \sim \frac{\ln n }{n} $ given $|\alpha_n| = o(\ln n)$. Then we further get $P_n = {K_n}^2 \big/\frac{{K_n}^2}{P_n} =  \Omega \big(\ln n \big/\frac{\ln n }{n} \hspace{1pt}\big) = \Omega (n)$ and
$\frac{K_n}{P_n} =  \frac{{K_n}^2}{P_n} \big/ K_n = O\big(\frac{\ln n }{n}\big/ \sqrt{\ln n} \hspace{1pt}\big) = o(1)$. Hence, as mentioned above, Lemma  \ref{lemma-1} is established.


Now we prove Lemma \ref{lemma-2}. We let $q_n$ be the edge probability in a uniform random intersection graph $G_u(n, P_n, K_n)$; i.e., two nodes in $G_u(n, P_n, K_n)$ have an edge in between with probability $q_n$. Under conditions of Lemma \ref{lemma-2}, given $|\alpha_n| = o(\ln n)$, we have
 \begin{align}
\frac{{K_n}^2}{P_n}  = \frac{\ln  n + {(k-1)} \ln \ln n +
 {\alpha_n}}{n} \sim \frac{\ln n }{n} . \label{PnKKst}
  \end{align}
  Hence, from \cite[Lemma 8-Property (a)]{ZhaoYaganGligor}, it follows that
 \begin{align}
 q_n = \frac{{K_n}^2}{P_n} \left[ 1 \pm O\bigg(\frac{{K_n}^2}{P_n}\bigg) \right] \sim \frac{\ln n }{n} . \label{eq_pe_lnnn}
 \end{align}
 Then, by \cite[Section 3]{mobihoca}, Lemma \ref{lemma-2} will follow once we show Lemma  \ref{lemma-3} below, where $\mathcal {V}_n = \{v_1, v_2,
\ldots, v_n \}$ is the set of nodes in graph $G_u(n, P_n, K_n)$.
\begin{lem}  \label{lemma-3}
For a uniform random intersection graph $G_u(n, P_n, K_n)$ under $K_n = \Omega \big(\sqrt{\ln n}\hspace{2pt}\big)$ and $ q_n \sim \frac{\ln n }{n}$, it follows for integers $m\geq 1$ and $h \geq 0$ that
\begin{align}
&  \mathbb{P} [\text{Nodes }v_{1}, v_{2}, \ldots, v_{m}\text{ have
degree }h] \nonumber \\
&   \quad \sim (h!)^{-m}  (n q_n)^{hm} e^{-m n q_n}.
\label{eqn_node_v12n}
\end{align}
\end{lem}

The rest of this section is devoted to proving Lemma  \ref{lemma-3}.


In a uniform random intersection graph $G_u(n, P_n, K_n)$, recalling that
 $\mathcal {V}_n = \{v_1, v_2,
\ldots, v_n \}$ is the set of nodes, we let $S_i$ be the set of $K_n$
distinct objects assigned to node $v_i \in \mathcal {V}_n$.
We further define $\mathcal {V}_m$ as $ \{v_1, v_2, \ldots, v_m\}$ and
$\overline{\mathcal {V}_m} $ as $ \mathcal {V}_n \setminus \mathcal {V}_m
$. Among nodes in $\overline{\mathcal {V}_m}$, we denote by $N_i$
the set of nodes neighboring to $v_i$ for $i=1,2,\ldots,m$. We
denote $N_i \bcap N_j$ by $N_{ij}$, and $S_i \bcap S_j$ by $S_{ij}$.

We have the following two observations:
\begin{itemize}
  \item [i)] If node $v_i$ has degree $h$,
then $|N_{i}| \leq h$, where the equal sign holds if and only if
$v_i$ is directly connected to none of nodes in $\mathcal{V}_m \setminus
\{v_i\}$; i.e., if and only if $\bigcap_{j \in \{1,2,\ldots,m\}\setminus\{i\}}
(S_{ij}=\emptyset)$ happens.  \vspace{2pt}
    \item [ii)] If $|N_{i}| \leq h$ for any $i=1,2,\ldots,m$, then
\begin{align}
& \bigg|\bigcup_{1\leq i \leq m} N_{i}\bigg| \leq \sum_{1\leq i \leq
m}N_{i}  \leq  hm , \label{Nileq}
\end{align}
where the two equal signs in (\ref{Nileq}) \emph{both} hold if and only if \vspace{-1pt}
\begin{align}
\bigg(\bigcap_{1\leq i <j \leq m} (N_{ij}=\emptyset)\bigg)  \bcap
\bigg(\bigcap_{1\leq i \leq m}(|N_{i}| = h)\bigg).  \vspace{-1pt}\label{Nij}
\end{align}
\end{itemize}

From i) and ii) above, if nodes $v_{1}, v_{2}, \ldots, v_{m}$ have degree $h$,
we have either of the following two cases:
\begin{itemize}
  \item [(a)] Any two of $v_{1}, v_{2}, \ldots, v_{m}$ have no edge in between (namely, $\bigcap_{1\leq i <j \leq m}
(S_{ij}=\emptyset)$); and event (\ref{Nij}) happens. \vspace{1pt}
  \item [(b)] $\big|\bigcup_{1\leq i \leq m} N_{i}\big|
  \leq hm -1$.
\end{itemize}
In addition, if case (a) happens, then nodes $v_{1}, v_{2}, \ldots,
v_{m}$ have degree $h$. However, if case (b) occurs, there is no
such conclusion. With $P_a$ (resp., $P_b$) denoting the probability
of case (a) (resp., case (b)), we obtain  \vspace{-1pt}
\begin{align}
 & P_a \leq \mathbb{P} [\text{Nodes }v_{1}, v_{2}, \ldots, v_{m}\text{ have
degree }h] \leq P_a + P_b,  \vspace{-1pt}\nonumber
\end{align}
where  \vspace{-1pt}
\begin{align}
P_a =  \mathbb{P}\bigg[ \bigg(\bigcap_{1\leq i <j \leq m}
(S_{ij}=\emptyset)\bigg)  &   \bcap \bigg(\bigcap_{1\leq i <j \leq m}
(N_{ij}=\emptyset)\bigg)   \nonumber \\
&    \bcap \bigg(\bigcap_{1\leq i \leq
m}(|N_{i}| = h)\bigg)\bigg],  \vspace{-1pt} \nonumber
\end{align}
and  \vspace{-1pt}
\begin{align}
 P_b & = \mathbb{P}\bigg[\bigg|\bigcup_{1\leq i \leq m} N_{i}\bigg|
  \leq hm -1\bigg].  \vspace{-1pt} \nonumber
\end{align}
Hence, (\ref{eqn_node_v12n}) holds after we prove the
following (\ref{prop2}) and (\ref{prop1}):
\begin{align}
 P_b & =   o \left((nq_n)^{hm} e^{-m n q_n}\right). \label{prop2}
\end{align}
and
\begin{align}
P_a & \sim (h!)^{-m} (n q_n)^{hm} e^{-m n q_n} \cdot
 [1+o(1)], \label{prop1}
\end{align}

We will prove (\ref{prop2}) and (\ref{prop1}) below.
 We let $\mathbb{S}_m$ denote the tuple $(S_1,S_2,\ldots,S_m)$. The
expression ``$\given \mathbb{S}_m = \mathbb{S}_m^*$'' means ``given
$S_1=S_1^*,S_2=S_2^*,\ldots,S_m=S_m^*$'', where $\mathbb{S}_m^* =
(S_1^*,S_2^*,\ldots,S_m^*)$ with $S_1^*,S_2^*,\ldots,S_m^*$ being
arbitrary $K_n$-size subsets of the object pool $\mathcal {P}_n$ (see Page \pageref{poolref} in the graph definition for the meaning of $\mathcal {P}_n$). Note that
$S_{ij}^{*} : = S_{i}^{*} \cap S_{j}^{*}$. For two different nodes $v$ and $w$ in $G_u(n, P_n, K_n)$, we use $v\leftrightarrow w$ to denote
the event that there is an edge between $v$ and $w$; i.e., the symbol ``$\leftrightarrow $'' means
``is directly connected with''.

\subsection{The Proof of (\ref{prop2})}

Let $w$ be an arbitrary node in $\overline{\mathcal{V}_m}$. The event $w \in \mathlarger{\cup}_{1\leq i \leq m} N_{i}$ means $w\leftrightarrow \text{ at least one of nodes in }\mathcal{V}_m$, which for different $w$ would be independent given $\mathbb{S}_m= \mathbb{S}_m^*$. Then we have
\begin{align}
  &  \mathbb{P}\bigg[\bigg|
\bigcup_{1\leq i \leq m} N_{i}\bigg| =  t \bgiven
\mathbb{S}_m= \mathbb{S}_m^*\bigg]  \label{t} \\
& = \frac{(n-m)!}{t!(n-m-t)!}   \nonumber \\
& \times \big\{\mathbb{P}[w\leftrightarrow \text{ at least one of nodes in }\mathcal{V}_m \given
\mathbb{S}_m= \mathbb{S}_m^*]\big\}^t  \nonumber \\
&   \times \big\{\mathbb{P}[w\leftrightarrow \text{none of
nodes in }\mathcal{V}_m \given \mathbb{S}_m= \mathbb{S}_m^*]\big\}^{n-m-t}.
\label{x}
\end{align}

By the union bound, it holds that
\begin{align}
& \mathbb{P}[w\leftrightarrow \text{at least one of
nodes in }\mathcal{V}_m \given \mathbb{S}_m= \mathbb{S}_m^*]  \nonumber \\
& \leq \sum_{1\leq i \leq m}\mathbb{P}
  [w\leftrightarrow v_i \given \mathbb{S}_m= \mathbb{S}_m^*] = m q_n,\label{lll}
\end{align}
which yields
\begin{align}
& \mathbb{P}[w\leftrightarrow \text{none of nodes in }\mathcal{V}_m \given
\mathbb{S}_m= \mathbb{S}_m^*] \geq  1 - m q_n. \label{sstar3}
\end{align}
In addition, we find
\begin{align}
& \mathbb{P}[w\leftrightarrow \text{none of nodes in }\mathcal{V}_m \given
\mathbb{S}_m= \mathbb{S}_m^*] \nonumber \\ &  = \frac{\binom{P_n -
|\bigcup_{1\leq i \leq m} S_i^*|}{K_n}}{\binom{P_n}{K_n}} \nonumber \\
& \leq (1-q_n)^{{K_n}^{-1}{|\bigcup_{1\leq i \leq m} S_i^*|}} \quad
\text{(by \cite[Lemma 5.1]{yagan_onoff})}  \nonumber \\
& \leq e^{-{K_n}^{-1}q_n{|\bigcup_{1\leq i \leq m} S_i^*|}} \quad
\text{(by $1+x \leq e^x$ for any real $x$)}.
\label{sstar4}
\end{align}

We will prove
\begin{align}
& \sum_{\mathbb{S}_m^*} \hspace{-2pt} \Bigg\{ \hspace{-1pt}\mathbb{P}[\mathbb{S}_m =
\mathbb{S}_m^*]  \hspace{-1pt} \times \hspace{-1pt}  \bigg\{ \hspace{-1pt} \mathbb{P}\bigg[\begin{array}{l}w\leftrightarrow \text{none of}\\ \text{nodes in }\mathcal{V}_m\end{array} \hspace{-3pt} \bigg|\hspace{2pt} \mathbb{S}_m= \mathbb{S}_m^*\bigg]
\hspace{-1pt}\bigg\}^{n-m-hm}
 \hspace{-1pt}\Bigg\} \label{sstar2} \\
  & \quad \leq  e^{-m
n q_n} \cdot
 [1+o(1)] . \label{sstar}
\end{align}
From (\ref{x}) (\ref{lll}) and (\ref{sstar}), we derive
\begin{align}
P_b  & = \mathbb{P}\bigg[\bigg|\bigcup_{1\leq i \leq m} N_{i}\bigg|
  \leq hm -1\bigg]   \nonumber \\ &= \sum_{t=0}^{hm -1}
   \sum_{\mathbb{S}_m^*} \Big\{ \mathbb{P}[\mathbb{S}_m = \mathbb{S}_m^*]
    \cdot (\ref{t})
 \Big\}
  \nonumber \\
  & \leq \sum_{t=0}^{hm -1} \Big[ n^t \cdot (m q_n)^t  \cdot  (\ref{sstar2})\Big]
 \nonumber \\ &   \leq (nq_n)^{hm} e^{-m n q_n} \cdot
 [1+o(1)] \cdot m^{hm}
  \sum_{t=0}^{hm -1} (mnq_n)^{t-hm}
 . \label{bp}
\end{align}
Applying (\ref{eq_pe_lnnn}) to (\ref{bp}), we obtain (\ref{prop2}). Hence,
we complete the proof of (\ref{prop2}) once showing (\ref{sstar}),
whose proof is detailed below.

From (\ref{eq_pe_lnnn}) (\ref{sstar3}) and (\ref{sstar4}), we have
\begin{align}
\textrm{(\ref{sstar2})} \hspace{-1pt}& \hspace{-1pt}\leq\hspace{-1pt} (1 - m q_n)^{-m-hm}  \nonumber  \\
  \hspace{-1pt}& \hspace{-1pt}\quad \times \sum_{\mathbb{S}_m^*}
\Big\{ \mathbb{P}[\mathbb{S}_m = \mathbb{S}_m^*] \cdot
e^{-{K_n}^{-1}nq_n{|\bigcup_{1\leq i \leq m} S_i^*|}}
\Big\} \nonumber  \\
  \hspace{-1pt}& \hspace{-1pt}\leq\hspace{-1pt}[1+o(1)]\hspace{-1pt}\cdot\hspace{-1pt}\sum_{\mathbb{S}_m^*}
\Big\{ \mathbb{P}[\mathbb{S}_m \hspace{-1pt}=\hspace{-1pt} \mathbb{S}_m^*] \hspace{-1pt}\cdot\hspace{-1pt}
e^{-{K_n}^{-1}nq_n{|\bigcup_{1\leq i \leq m} S_i^*|}} \Big\},
\end{align}
so (\ref{sstar}) holds once we demonstrate
\begin{align}
& \sum_{\mathbb{S}_m^*} \Big\{ \mathbb{P}[\mathbb{S}_m =
\mathbb{S}_m^*]  \cdot e^{-{K_n}^{-1}nq_n{|\bigcup_{1\leq i \leq m}
S_i^*|}} \Big\}  \nonumber  \\
  & \quad \leq e^{-m n q_n} \cdot
 [1+o(1)] . \label{ms}
\end{align}
We denote the left hand side of (\ref{ms}) by $Z_{m,n}$. Dividing $\mathbb{S}_{m}^*$ into two parts $\mathbb{S}_{m-1}^*$
and $S_m^*$,
 we derive
\begin{align}
  Z_{m,n}
  &=  \sum_{\begin{subarray}{c}\mathbb{S}_{m-1}^*
  \\S_m^*\end{subarray}} \Big\{ \mathbb{P}[(\mathbb{S}_{m-1} = \mathbb{S}_{m-1}^*)
  \bcap(S_m = S_m^*)]  \nonumber  \\
  & \quad\quad\quad\quad \times e^{-{K_n}^{-1}nq_n{|\bigcup_{1\leq i \leq m}
S_i^*|}} \Big\}\nonumber \\  &= \sum_{\mathbb{S}_{m-1}^*}
\mathbb{P}[\mathbb{S}_{m-1} = \mathbb{S}_{m-1}^*] \bigg\{
e^{-{K_n}^{-1}nq_n{|\bigcup_{1\leq i \leq m-1} S_i^*|}} \nonumber \\
&  \quad\quad \times \sum_{S_m^* }
\mathbb{P}[ S_m = S_m^* ] e^{-{K_n}^{-1}nq_n{|S_m^* \setminus
\bigcup_{1\leq i \leq m-1} S_i^*|}}\bigg\} ,\label{HnmHnm1}
\end{align}
 where
\begin{align}
&  \sum_{S_m^* } \mathbb{P}[ S_m = S_m^* ] e^{-{K_n}^{-1}nq_n{|S_m^*
\setminus \bigcup_{1\leq i \leq m-1} S_i^*|}} \nonumber \\ &  \leq
e^{-n q_n}\sum_{S_m^* } \mathbb{P}[ S_m = S_m^* ] e^{{K_n}^{-1}{ n
q_n}\big|S_m^*  \cap
 \big(\bigcup_{i =1}^{m-1} S_{i }^* \big)  \big|}  \nonumber \\ &
 =  e^{-n q_n} \sum_{r=0}^{K_n}
\mathbb{P}\bigg[\bigg|S_m\bcap \bigg(\bigcup_{i =1}^{m-1}S_{i
}^*\bigg)\bigg| = r \bigg] e^{{K_n}^{-1}{n q_nr} } . \label{SS}
\end{align}

Denoting $\big|\bigcup_{i=1}^{m-1}S_{i}^*\big|$ by $v$, then for $r$ satisfying the conditions
$0 \leq r \leq |S_m^*| = K_n$ and $S_m^* \bcup
\big(\bigcup_{i=1}^{m-1}S_{i}^*\big) = K_n + v - r \leq P_n $ (i.e., for $r \in [\max\{0, K_n + v - P_n\}  , K_n] $), we
obtain
\begin{align}
 \hspace{-5pt} \mathbb{P}\bigg[\bigg|S_m \hspace{-1pt} \bcap  \hspace{-1pt}
\bigg(\bigcup_{i=1}^{m-1}S_{i}^*\bigg)\bigg|  \hspace{-1pt} =  \hspace{-1pt} r \bigg] &  =
 \binom{v}{r} \binom{P_n - v}{K_n - r} \bigg/{\binom{P_n}{K_n}},
\label{probsm}
\end{align}
which together with $ K_n \leq  v \leq m K_n$ yields
\begin{align}
 & \text{L.H.S. of (\ref{probsm})} \nonumber \\ & \quad \leq \frac{(m K_n)^r}{r!}   \cdot
 \frac{(P_n - K_n)^{K_n - r}}{(K_n - r)!}  \cdot  \frac{K_n !}{(P_n - K_n)^{K_n}}\nonumber
\\&  \quad \leq \frac{1}{r!} \bigg( \frac{m {K_n}^2}{P_n - K_n}\bigg)^r \text{ for $r \in [\max\{0, K_n + v - P_n\}  , K_n] $}. \label{probsm2}
\end{align}
Also, it is clear that
\begin{align}
 \text{L.H.S. of (\ref{probsm})} = 0 \text{ for $r \notin [\max\{0, K_n + v - P_n\}  , K_n] $} .\label{probsm20clr}
\end{align}

Applying (\ref{probsm2}) and (\ref{probsm20clr}) to (\ref{SS}), we establish
\begin{align}
&  \sum_{S_m^* } \mathbb{P}[ S_m = S_m^* ] e^{-{K_n}^{-1}nq_n{|S_m^*
\setminus \bigcup_{1\leq i \leq m-1} S_i^*|}} \nonumber \\ &  \leq
 e^{-n q_n} \sum_{r=0}^{K_n} \frac{1}{r!} \bigg( \frac{m {K_n}^2}{P_n - K_n}\bigg)^r
  e^{{K_n}^{-1}{n q_nr} }  \nonumber \\ &
 \leq  e^{-n q_n} \cdot e^{\frac{m {K_n}^2}{P_n - K_n}
 \cdot e^{{K_n}^{-1}{n q_n}}} . \label{psnm}
\end{align}
From $K_n = \Omega \big(\sqrt{\ln n}\hspace{2pt}\big)$ and (\ref{PnKKst}) (i.e., $\frac{{K_n}^2}{P_n}\sim \frac{\ln n}{n}$), we have $P_n = \omega(K_n)$ and further
\begin{align}
\frac{m {K_n}^2}{P_n - K_n} &  \sim \frac{m{K_n}^2}{P_n} \sim
\frac{m\ln n}{n}. \label{e1}
\end{align}
For an arbitrary $\epsilon > 0$, from (\ref{eq_pe_lnnn}), we obtain
$q_n \leq (1+\epsilon)\frac{\ln n}{n}$ \vspace{1pt} for all $n$ sufficiently
large, which with   
$K_n = \Omega \big(\sqrt{\ln n}\hspace{2pt}\big) \geq 2$ yields that for all $n$ sufficiently large,
\begin{align}
e^{{K_n}^{-1}{n q_n}} &  \leq e^{\frac{1}{2}(1+\epsilon)\ln n} =
n^{\frac{1}{2}(1+\epsilon)}.  \label{e2}
\end{align}
From (\ref{e1}) and (\ref{e2}), we get
\begin{align}
\frac{m {K_n}^2}{P_n - K_n} \cdot e^{{K_n}^{-1}{n q_n}}  &  
 \leq m\ln n \cdot n^{\frac{1}{2}(\epsilon-1)}  \cdot [1+o(1)] .
\label{e3}
\end{align}
Since $\epsilon > 0$ is arbitrary, it follows from (\ref{e3}) that
for arbitrary $0<c<\frac{1}{2}$, then for all $n$ sufficiently
large, it is clear that
\begin{align}
\frac{m {K_n}^2}{P_n - K_n} \cdot e^{{K_n}^{-1}{n q_n}} & \leq
n^{-c}. \label{e4}
\end{align}
Using (\ref{e4}) in (\ref{psnm}), for all $n$ sufficiently large, it follows that
\begin{align}
 \sum_{S_m^* } \mathbb{P}[ S_m = S_m^* ] e^{-{K_n}^{-1}nq_n{|S_m^*
\setminus \bigcup_{1\leq i \leq m-1} S_i^*|}} &  \leq e^{-n q_n}
\cdot e^{n^{-c}} .  \label{e5}
\end{align}
Substituting (\ref{e5}) into (\ref{HnmHnm1}), for all $n$
sufficiently large, we obtain
\begin{align}
 &\hspace{-2pt} Z_{m,n}  \nonumber  \\
 & \hspace{-3pt}\leq\hspace{-1.5pt} e^{-n q_n} \hspace{-1.5pt}\cdot\hspace{-1.5pt} e^{n^{-c}}
\hspace{-1.5pt}\cdot \hspace{-1.5pt}\sum_{\mathbb{S}_{m-1}^*} \hspace{-1pt}\mathbb{P}[\mathbb{S}_{m-1} \hspace{-1.5pt}=\hspace{-1.5pt}
\mathbb{S}_{m-1}^*]
e^{-{K_n}^{-1}nq_n{|\bigcup_{1\leq i \leq m-1} S_i^*|}} \nonumber \\
&  \hspace{-3pt}\leq\hspace{-1pt} e^{-n q_n} \hspace{-1pt}\cdot \hspace{-1pt}e^{n^{-c}} \hspace{-1pt}\cdot\hspace{-1pt} Z_{m-1,n}.
\end{align}
We then evaluate $Z_{2,n}$. By (\ref{ms}), it holds that
\begin{align}
& \hspace{-2pt}Z_{2,n}  \nonumber  \\
  &  \hspace{-3pt}=\hspace{-1.5pt}\sum_{S_1^*}\hspace{-1pt}\sum_{S_2^*} \hspace{-1pt}\Big\{ \hspace{-1pt}\mathbb{P}[(S_1  \hspace{-1pt}=\hspace{-1pt}
S_1^*)\hspace{-1pt}\bcap \hspace{-1pt}(S_2  \hspace{-1pt}=\hspace{-1pt} S_2^*)] \hspace{-1.5pt}\cdot\hspace{-1.5pt}e^{-{K_n}^{-1}nq_n{|S_1^* \bcup
S_2^*|}} \Big\}\nonumber \\  & = \sum_{S_1^*} \mathbb{P}[ S_1  =
S_1^* ] \sum_{S_2^*} \mathbb{P}[ S_2  = S_2^* ]
e^{-{K_n}^{-1}nq_n{|S_1^* \bcup S_2^*|}}. \label{mm1}
\end{align}
Setting $m=2$ in (\ref{e5}), for all $n$ sufficiently large, we derive
\begin{align}
\sum_{S_2^*} \mathbb{P}[ S_2  = S_2^* ] e^{-{K_n}^{-1}nq_n{| S_2^*
\setminus S_1^*|}} & \leq e^{-n q_n} \cdot e^{n^{-c}} . \nonumber
\end{align}
Then for all $n$ sufficiently large, it follows that
\begin{align}
& \sum_{S_2^*} \mathbb{P}[ S_2  = S_2^* ] e^{-{K_n}^{-1}nq_n{|S_1^*
\bcup S_2^*|}}  \nonumber  \\
  &  =  e^{-n q_n} \sum_{S_2^*} \mathbb{P}[ S_2  =
S_2^* ] e^{-{K_n}^{-1}nq_n{| S_2^* \setminus S_1^*|}}  \nonumber \\
& \leq  e^{-2n q_n} \cdot e^{n^{-c}} . \label{e6}
\end{align}

From (\ref{mm1}) and (\ref{e6}), for all $n$ sufficiently large, we
obtain
\begin{align}
  Z_{m,n} & \leq \big(e^{-n q_n} \cdot e^{n^{-c}}\big)^{m-2}
\cdot Z_{2,n} \nonumber \\
&  \leq \big(e^{-n q_n} \cdot e^{n^{-c}}\big)^{m-2} \cdot e^{-2n
q_n} \cdot e^{n^{-c}}
\nonumber \\
&  \leq e^{-mn q_n} \cdot e^{(m-1)n^{-c}} .
\end{align}
Letting $n \to \infty$, we finally establish
\begin{align}
  Z_{m,n} & \leq e^{-m n q_n} \cdot
 [1+o(1)] ; \nonumber
\end{align}
i.e., (\ref{ms}) is proved. As explained, (\ref{sstar})
 and then (\ref{prop2}) follow.

\subsection{The Proof of (\ref{prop1})}

Again let $w$ be an arbitrary node in $\overline{\mathcal{V}_m}$. We have
\begin{align}
& \mathbb{P}\bigg[\bigg(\bigcap_{1\leq i <j \leq m}
(N_{ij}=\emptyset)\bigg)  \bcap \bigg(\bigcap_{1\leq i \leq
m}(|N_{i}| = h)\bigg) \bgiven \mathbb{S}_m= \mathbb{S}_m^*\bigg]
 \label{h} \\ & = \frac{(n-m)!}{(h!)^m(n-m-hm)!}  \nonumber \\
  & \quad \times \prod_{1\leq i \leq m}\left(\left\{\mathbb{P}
  \left[\begin{array}{l}w\leftrightarrow v_i,\\\text{but }w\leftrightarrow\text{none of}\\\text{nodes
in }\mathcal{V}_m \setminus \{v_i\}\end{array}\Bigg|\hspace{3pt} \mathbb{S}_m= \mathbb{S}_m^*\right]\right\}^h\right)  \nonumber \\
& \quad \times \big\{\mathbb{P}[w\leftrightarrow \text{none of
nodes in }\mathcal{V}_m \given \mathbb{S}_m= \mathbb{S}_m^*]\big\}^{n-m-hm}
\label{y}
\end{align}
and
\begin{align}
P_a  & = \sum_{\mathbb{S}_m^*:\hspace{2pt}\bigcap_{1\leq i <j \leq
m} (S_{ij}^*=\emptyset)} \Big\{ \mathbb{P}[\mathbb{S}_m =
\mathbb{S}_m^*] \cdot (\ref{h})
 \Big\} \label{pra} ,
\end{align}
where $S_{ij}^{*} : = S_{i}^{*} \cap S_{j}^{*}$.

For $i=1,2,\ldots,m$, under
$\mathbb{S}_m^*:\hspace{2pt}\bigcap_{1\leq i <j \leq m}
(S_{ij}^*=\emptyset)$, it follows that
\begin{align}
& \mathbb{P}
  [w\leftrightarrow v_i,\text{ but none of nodes
in }\mathcal{V}_m \setminus \{v_i\}\given \mathbb{S}_m= \mathbb{S}_m^*]  \nonumber \\
& \geq \mathbb{P}
  [w\leftrightarrow v_i \given \mathbb{S}_m= \mathbb{S}_m^*]
 \nonumber \\
&  \quad - \sum_{\begin{subarray}{c} 1 \leq j \leq m \\ j\neq
i\end{subarray}} \mathbb{P}
  [w\leftrightarrow \text{both }v_i\text{ and }v_j
  \given \mathbb{S}_m= \mathbb{S}_m^*] , \label{ps2-nctacsdf}
\end{align}
where we note
\begin{align}
 \mathbb{P} [w\leftrightarrow v_i \given \mathbb{S}_m= \mathbb{S}_m^*] = q_n,  \label{nsw0}
\end{align}
and
\begin{align}
& \mathbb{P} [w\leftrightarrow \text{both }v_i\text{ and }v_j
  \given \mathbb{S}_m= \mathbb{S}_m^*]
\nonumber \\ &   =  \mathbb{P}
  [w\leftrightarrow v_i \given \mathbb{S}_m= \mathbb{S}_m^*]
  + \mathbb{P}
  [w\leftrightarrow v_j \given \mathbb{S}_m= \mathbb{S}_m^*]
 \nonumber \\ & \quad - \mathbb{P}
  [ (w\leftrightarrow v_i) \bcup (w\leftrightarrow v_j) \given \mathbb{S}_m= \mathbb{S}_m^*]
  \nonumber \\ &   = q_n + q_n -  \binom{P_n - 2K_n}{K_n} \bigg/   \binom{P_n}{K_n} \label{nsw1}
\end{align}
given $\mathbb{S}_m^*:\hspace{2pt}\bigcap_{1\leq i <j \leq m}
(S_{ij}^*=\emptyset)$.
 From \cite[Lemma 5.1]{yagan_onoff}, we get $\binom{P_n - 2K_n}{K_n} \big/   \binom{P_n}{K_n} \leq ( 1- {q_n })^2$, which with (\ref{nsw0}) and (\ref{nsw1}) are used in (\ref{ps2-nctacsdf}) to derive
\begin{align}
& \mathbb{P}
  [w\leftrightarrow v_i,\text{ but none of nodes
in }\mathcal{V}_m \setminus \{v_i\}\given \mathbb{S}_m= \mathbb{S}_m^*]  \nonumber \\
& \quad \geq  q_n - (m-1) \cdot 2{q_n} ^2  . \label{ps2}
\end{align}

Substituting (\ref{sstar3}) and (\ref{ps2}) to (\ref{y}), and then
from (\ref{pra}), we obtain
\begin{align}
P_a  & \geq \frac{(n-m-hm)^{hm}}{(h!)^m} \cdot [q_n - 2(m-1) q_n
^2]^{hm}    \nonumber \\
&  \quad   \times (1-mq_n)^{n-m-hm} \sum_{\mathbb{S}_m^*:\hspace{2pt}\bigcap_{1\leq i
<j \leq m} (S_{ij}^*=\emptyset)}  \mathbb{P}[\mathbb{S}_m =
\mathbb{S}_m^*]. \nonumber
\end{align}
Then from (\ref{eq_pe_lnnn}), it further hold that
\begin{align}
P_a  & \geq \frac{n^{hm}}{(h!)^m} \cdot (q_n)^{hm}
  \cdot  e^{- m n q_n}  \nonumber \\
&  \quad \times [1-o(1)] \cdot
  \mathbb{P}\bigg[\bigcap_{1\leq i <j \leq m} (S_{ij}=\emptyset)\bigg]
    . \label{poa1}
\end{align}

From (\ref{sstar4}), under
$\mathbb{S}_m^*:\hspace{2pt}\bigcap_{1\leq i <j \leq m}
(S_{ij}^*=\emptyset)$, it holds that
\begin{align}
 \mathbb{P}[w\leftrightarrow \text{none of nodes in }\mathcal{V}_m \given
\mathbb{S}_m= \mathbb{S}_m^*] & \leq e^{- m q_n} . \label{ps3a}
\end{align}

For each $i=1,2,\ldots,m$, we have
\begin{align}
& \mathbb{P}
  [w\leftrightarrow v_i,\text{ but }w\leftrightarrow \text{none of nodes
in }\mathcal{V}_m \setminus \{v_i\}\given \mathbb{S}_m= \mathbb{S}_m^*]  \nonumber \\
& \leq \mathbb{P}
  [w\leftrightarrow v_i \given \mathbb{S}_m= \mathbb{S}_m^*] =
  q_n. \label{ps}
\end{align}
Substituting (\ref{ps}) and (\ref{ps3a}) to (\ref{y}), and then from
(\ref{pra}), we obtain
\begin{align}
P_a  & \hspace{-1.5pt}\leq\hspace{-1.5pt} \frac{n^{hm}}{(h!)^m} \hspace{-1.5pt}\cdot\hspace{-1.5pt} (q_n)^{hm}
  \hspace{-1.5pt}\cdot\hspace{-1.5pt}  e^{- m n q_n} \hspace{-1.5pt}\cdot\hspace{-1.5pt} \sum_{\mathbb{S}_m^*:\hspace{2pt}\hspace{-1.5pt}\bigcap_{1\leq i
<j \leq m} (S_{ij}^*=\emptyset)}  \hspace{-1.5pt}\mathbb{P}[\mathbb{S}_m \hspace{-1.5pt}=\hspace{-1.5pt}
\mathbb{S}_m^*] \nonumber \\
&   =  \frac{n^{hm}}{(h!)^m} \cdot (q_n)^{hm}
  \cdot  e^{- m n q_n} \cdot
  \mathbb{P}\bigg[\bigcap_{1\leq i <j \leq m} (S_{ij}=\emptyset)\bigg]
 .  \label{poa2}
\end{align}

From (\ref{poa1}) and (\ref{poa2}),  we obtain\vspace{-1pt}
\begin{align}
P_a  &   \sim \frac{n^{hm}}{(h!)^m} \cdot (q_n)^{hm}
  \cdot  e^{- m n q_n} \cdot
  \mathbb{P}\bigg[\bigcap_{1\leq i <j \leq m} (S_{ij}=\emptyset)\bigg].
  \label{y2} \vspace{-1pt}
\end{align}
By the union bound, it is clear that\vspace{-1pt}
\begin{align}
& \mathbb{P}\bigg[\bigcap_{1\leq i <j \leq m} (S_{ij}=\emptyset)\bigg]\vspace{-1pt}
\nonumber \\  &    \geq 1 - \sum_{1\leq i <j \leq m}
\mathbb{P}[S_{ij}\neq \emptyset] = 1 - \binom{m}{2}q_n. \label{m2ps}\vspace{-1pt}
\end{align}
From (\ref{eq_pe_lnnn}) and (\ref{m2ps}), since a
probability is at most $1$, we get\vspace{-1pt}
\begin{align}
\lim_{n \to \infty}\mathbb{P}\bigg[\bigcap_{1\leq i <j \leq m}
(S_{ij}=\emptyset)\bigg] & = 1 .\vspace{-1pt} \label{m2ps2}
\end{align}
Using (\ref{m2ps2}) in (\ref{y2}), we establish (\ref{prop1}).

\section{Establishing Theorems \ref{thm:rig:rb}--\ref{thm:grig:rb}}
\label{sec:thmprf:krb}
 Theorems \ref{thm:rig:rb}--\ref{thm:grig:rb} present results on
$k$-robustness for binomial/uniform/general random intersection graphs. We prove
Theorems \ref{thm:rig:rb}--\ref{thm:grig:rb} in this section and start with explaining the idea below. First, the zero-law of Theorem  \ref{thm:rig:rb} is established from the zero-law of Theorem  \ref{thm:rig} since $k$-robustness implies the property of minimum node degree being at least $k$ from Lemma \ref{lem-k-robu-mnd} above, while the one-law of Theorem  \ref{thm:rig:rb} is proven
from the coupling between binomial random intersection graphs and Erd\H{o}s--R\'{e}nyi graphs given by Lemma \ref{cp_rig_er} of Section \ref{secfagcp_rig_er}. Second, the zero-law of Theorem  \ref{thm:urig:rb} is demonstrated from the zero-law of Theorem  \ref{thm:urig} because $k$-robustness implies the property of minimum node degree being at least $k$ from Lemma \ref{lem-k-robu-mnd} above, while the one-law of Theorem  \ref{thm:urig:rb} is established
from the coupling between binomial random intersection graphs and uniform random intersection graphs given by Lemma \ref{rkgikg} of Section \ref{secacp_rkgikg}. Finally, both the zero-law and one-law of Theorem  \ref{thm:grig:rb} are proved from the coupling between general random intersection graphs and uniform random intersection graphs given by Lemma \ref{lem:cp} of Section \ref{secffalem:cp}.

\subsection{The Proof of Theorem \ref{thm:rig:rb}} \label{prf:thm:rig:rb}

Since $k$-robustness implies the property of minimum node degree being at least $k$ from Lemma \ref{lem-k-robu-mnd}, the zero-law of Theorem \ref{thm:rig:rb} is clear from
(\ref{uni-kon-0-mnd}) of Theorem \ref{thm:rig} in view that under conditions of Theorem \ref{thm:rig:rb},
  if $\lim_{n \to \infty}{\alpha_n}
=-\infty$, \begin{align}
 & \mathbb{P} \big[\hspace{2pt}G_b(n,P_n,p_n)\text{
is $k$-robust}.\hspace{2pt}\big] \nonumber \\
& \leq \mathbb{P} \bigg[\hspace{-2pt}\begin{array}{l}G_b(n,P_n,p_n)\text{
has a}\\\text{minimum node degree at least $k$}.\end{array}\hspace{-2pt}\bigg] \to 0,\text{ as }n \to \infty.
\label{prf:thm:rig:rb1}
 \end{align}
Note that Theorem \ref{thm:rig} uses $P_n = \omega
\big(n(\ln n)^5\big)$ while Theorem \ref{thm:rig:rb} uses $P_n = \Omega
\big(n(\ln n)^5\big)$. Above we can use (\ref{uni-kon-0-mnd}) since (\ref{uni-kon-0-mnd}) still holds under $P_n = \Omega
\big(n(\ln n)^5\big)$ as given in Remark \ref{rm} after  Theorem \ref{thm:rig}.

Below we prove the one-law of Theorem \ref{thm:rig:rb}. As explained in Appendix \ref{seca:conf:bin}, we can introduce an extra condition $|\alpha_n| = o(\ln n)$ in proving Theorem \ref{thm:rig:rb}. 
Given (\ref{thm:rig:pe:rb}) and $|\alpha_n| =
o(\ln n)$, we have
\begin{align}
 {p_n}^2 P_n & \sim
 \frac{\ln  n}{n} , \nonumber
\end{align}
%
which together with
condition $P_n = \Omega \big(n(\ln n)^5\big)$ leads to
\begin{align}
p_n & \sim
 \sqrt{\frac{\ln  n}{nP_n}} = O\Bigg(\sqrt{\frac{\ln  n}{n^2(\ln n)^5}}\hspace{2pt}\Bigg)
 = O\bigg(\frac{1}{n(\ln n)^2}\bigg).   \label{thm:rig:pnx}
\end{align}
Noting that (\ref{thm:rig:pnx}) implies condition $p_n = O\left( \frac{1}{n\ln n} \right)$ in Lemma \ref{cp_rig_er}, we apply Lemmas \ref{er_robust}, \ref{mono-gcp} and \ref{cp_rig_er}, and
condition (\ref{thm:rig:pe:rb}) to derive the following: there exists
$\hat{p}_n = \frac{\ln  n + {(k-1)} \ln \ln n + {\alpha_n} -
O(1)}{n}$ such that if $\lim_{n \to \infty}{\alpha_n} = \infty$,
\begin{align}
 & \mathbb{P}[\hspace{2pt}\text{Graph }G_b(n,P_n,p_n)
  \text{ is $k$-robust.}
\hspace{2pt}] \nonumber
\\ &  \geq \hspace{-1pt}
 \mathbb{P}[\hspace{1.5pt}\text{Graph }G(n,\hat{p}_n)
 \text{ is $k$-robust.}\hspace{1.5pt} ]  \hspace{-1pt}- \hspace{-1pt} o(1)
  \hspace{-1pt}\to \hspace{-1pt} 1,\text{ as }n  \hspace{-1pt}\to \hspace{-1pt} \infty. \label{prf:thm:rig:rb2}
 \end{align}
The proof of Theorem \ref{thm:rig:rb} is completed via
(\ref{prf:thm:rig:rb1}) and (\ref{prf:thm:rig:rb2}).  


\subsection{The Proof of Theorem \ref{thm:urig:rb}} \label{prf:thm:urig:rb}

As explained in Appendix \ref{seca:conf:unig}, we can introduce an extra condition $|\alpha_n| = o(\ln n)$ in proving Theorem \ref{thm:urig:rb}. Since $k$-robustness implies that the minimum node degree is at least $k$ from Lemma \ref{lem-k-robu-mnd},
   the zero-law of Theorem \ref{thm:urig:rb} is clear from Lemma \ref{lemma-2} in view that under conditions of Theorem \ref{thm:urig:rb} with the extra condition $|\alpha_n| = o(\ln n)$, if $\lim_{n \to \infty}{\alpha_n}
=-\infty$,
\begin{align}
 & \mathbb{P} \big[\hspace{2pt}G_u(n,P_n,K_n)\text{
is $k$-robust}.\hspace{2pt}\big] \nonumber \\
& \leq \mathbb{P} \bigg[\hspace{-2pt}\begin{array}{l}G_u(n,P_n,K_n)\text{
has a}\\\text{minimum node degree at least $k$}.\end{array}\hspace{-2pt}\bigg]\to 0,\text{ as }n \to \infty.
\label{prf:thm:urig:rb1}
 \end{align}

Below we establish the one-law of Theorem \ref{thm:urig:rb} with the
help of Theorem \ref{thm:rig:rb}.  Given $K_n = \Omega \big((\ln
n)^3\big) = \omega\left( \ln n \right)$, we use Lemma
\ref{cp_urig_rig} to obtain that with $p_n$ set by
\begin{align}
p_n & =  \frac{K_n}{P_n}
 \left(1 - \sqrt{\frac{3\ln
n}{K_n }}\hspace{2pt}\right), \label{pnexpr}
 \end{align}
it holds that
\begin{align}
 & \mathbb{P}[\hspace{2pt}\text{Graph }G_u(n,P_n,K_n)
  \text{ is $k$-robust.}
\hspace{2pt}] \nonumber
\\ & \quad \geq
 \mathbb{P}[\hspace{2pt}\text{Graph }G_b(n,P_n,p_n)
 \text{ is $k$-robust.}\hspace{2pt} ] - o(1). \label{robustcomp}
 \end{align}

From (\ref{thm:urig:pe:rb}) and $|\alpha_n| =
o(\ln n)$, we obtain $\frac{{K_n}^2}{P_n}   \sim \frac{\ln n}{n}$,
%
 which together with $K_n = \Omega \big((\ln n)^3\big) $ results in
\begin{align}
P_n & \sim \frac{n{K_n}^2}{\ln n} = \Omega \big(n(\ln n)^5\big),
\label{Pnlnn5}
\end{align}
From $K_n = \Omega \big((\ln n)^3\big) $ and (\ref{pnexpr}), it
follows that
\begin{align}
{p_n}^2 P_n & =  \left[\frac{K_n}{P_n}
 \left(1 - \sqrt{\frac{3\ln
n}{K_n }}\hspace{2pt}\right)\right]^2 \cdot P_n \nonumber \\ & =
\frac{{K_n}^2}{P_n}
 \cdot \left[1 -
 O\left(\frac{1}{ \ln n}\right)\right] .\label{pnPnlnn}
 \end{align}
By (\ref{thm:urig:pe:rb}) and (\ref{pnPnlnn}),
 it is clear that
\begin{align}
  {p_n}^2 P_n & =
 \frac{\ln  n + {(k-1)} \ln \ln n + {\alpha_n} - O(1)}{n}. \label{thm:rig:pe:rbcdx}
\end{align}
Given (\ref{Pnlnn5}) (\ref{thm:rig:pe:rbcdx}) and $|\alpha_n| =
o(\ln n)$, we use Theorem \ref{thm:rig:rb} and (\ref{robustcomp}) to
get that if $\lim_{n \to \infty}{\alpha_n} =\infty$, then
\begin{align}
 & \mathbb{P} \big[\hspace{2pt}G_u(n,P_n,K_n)\text{
is $k$-robust}.\hspace{2pt}\big] \to 1,\text{ as }n \to \infty.
\label{prf:thm:urig:rb2}
 \end{align}
The proof of Theorem \ref{thm:urig:rb} is completed via
(\ref{prf:thm:urig:rb1}) and (\ref{prf:thm:urig:rb2}).  



\subsection{The Proof of Theorem \ref{thm:grig:rb}} \label{prf:thm:grig:rb}
%

Similar to the process of proving Theorem \ref{thm:grig} with the
help of Theorem \ref{thm:urig}, we demonstrate Theorem \ref{thm:grig:rb} using Theorem \ref{thm:urig:rb}, which has been proved above.

 %

Given Lemmas \ref{mono-gcp} and \ref{lem:cp} and the fact that $k$-robustness is a monotone increasing graph property,
 we will show Theorem \ref{thm:grig:rb} 
 once 
 proving  for any $\epsilon_n = o\left(\frac{1}{\ln  n}\right)$ that
\begin{align}
 & \lim_{n \to \infty}\mathbb{P} \big[ G_u(n, P_n,
 (1 \pm \epsilon_n)\mathbb{E}[X_n])
 \text{
is $k$-robust}. \big] \nonumber  \\
&   =
\begin{cases} 0, &\text{ if $\lim_{n \to \infty}{\alpha_n}
=-\infty$}, \\  1, &\text{ if $\lim_{n \to \infty}{\alpha_n}
=\infty$.}   \end{cases} \label{kconn:rb}
 \end{align}
Under $\mathbb{E}[X_n] = \Omega \big(\sqrt{\ln n}\hspace{2pt}\big)$ and
$\epsilon_n = o\left(\frac{1}{\ln n}\right)$, it follows that $(1 \pm \epsilon_n)\mathbb{E}[X_n]
 =  \Omega \big(\sqrt{\ln n}\hspace{2pt}\big)$. From Theorem \ref{thm:urig:rb}, we will have (\ref{kconn}) once we prove that sequences $\gamma_n^{+}$ and $\gamma_n^{-}$ defined through
 \begin{align}
 \frac{\big\{(1\pm \epsilon_n)\mathbb{E}[X_n]\big\}^2}{P_n}
   & = \frac{\ln  n + {(k-1)}
 \ln \ln n + {\gamma_n^{\pm}}  }{n}\label{pe_epsilon4tac:rb}
 \end{align}
 satisfy
  \begin{align}
\lim_{n \to \infty} \gamma_n^{\pm} &   =
\begin{cases} -\infty, &\text{ if $\lim_{n \to \infty}{\alpha_n}
=-\infty$}, \\  \infty, &\text{ if $\lim_{n \to \infty}{\alpha_n}
=\infty$.}     \end{cases} \label{pe_epsilon4tac2:rb}
 \end{align}
Note that (\ref{thm:grig:pe:rb})  and (\ref{pe_epsilon4tac:rb})    are exactly the same
as (\ref{thm:grig:pe}) and (\ref{pe_epsilon4tac}), while  (\ref{pe_epsilon4tac2:rb}) is a subset of (\ref{pe_epsilon4tac2}). Since
(\ref{pe_epsilon4tac2tc}) follows from (\ref{thm:grig:pe})   (\ref{pe_epsilon4tac}) and $\epsilon_n = o\left(\frac{1}{\ln  n}\right)$,   we use (\ref{thm:grig:pe:rb})   (\ref{pe_epsilon4tac:rb}) and $\epsilon_n = o\left(\frac{1}{\ln  n}\right)$ to obtain (\ref{pe_epsilon4tac2tc}), which further yields (\ref{pe_epsilon4tac2:rb}). Therefore, as mentioned above,
we establish (\ref{kconn:rb}) and finally  Theorem \ref{thm:grig:rb}.

\section{Establishing Lemmas in Section \ref{sec:factlem}} \label{sec:prf:fact:lem}

Lemmas \ref{er_robust} and \ref{rkgikg} are clear in Section \ref{sec:factlem}. Below we prove Lemmas \ref{lem:cp}, \ref{cp_rig_er} and \ref{cp_urig_rig}.

 \subsection{The Proof of Lemma \ref{lem:cp}} \label{Couplinggeneraluniform}

According to \cite[Lemma 3]{Rybarczyk}, for any monotone increasing
graph property $\mathcal {I}$ and any $|\epsilon_n|<1$,
\begin{align}
 & \mathbb{P} \big[ G(n,P_n,\mathcal {D}_n)\text{
has $\mathcal {I}$}. \big]   -  \mathbb{P} \hspace{-1pt} \big[
G_u(n,\hspace{-1.3pt}P_n,\hspace{-1.3pt}(1\hspace{-1.3pt}-\hspace{-1.3pt}
\epsilon_n)\mathbb{E}[X_n])\text{\hspace{-.2pt} has $\mathcal
{I}$}.\hspace{-.2pt} \big] \nonumber \\
&  \geq  \big\{ 1 - \mathbb{P}\hspace{-.2pt}[X_n \hspace{-1.3pt} <
\hspace{-1.3pt}
(1\hspace{-1.3pt}-\hspace{-1.3pt}\epsilon_n)\mathbb{E}[X_n]] \big\}^n
- 1, \label{coupling1}
 \end{align}
and
\begin{align}
 &  \mathbb{P} \big[ G(n,P_n,\mathcal {D}_n)\text{
has $\mathcal {I}$}. \big]  - \mathbb{P} \hspace{-1pt} \big[
G_u(n,\hspace{-1.3pt}P_n,\hspace{-1.3pt}(1\hspace{-1.3pt}+\hspace{-1.3pt}
\epsilon_n)\mathbb{E}[X_n])\text{\hspace{-.2pt} has $\mathcal
{I}$}.\hspace{-.2pt} \big] \nonumber \\
&  \leq 1 -  \big\{1 - \mathbb{P}\hspace{-.2pt}[X_n \hspace{-1.3pt}
> \hspace{-1.3pt}
(1\hspace{-1.3pt}+\hspace{-1.3pt}\epsilon_n)\mathbb{E}[X_n]] \big\}^n.
\label{coupling2}
 \end{align}

 By (\ref{coupling1})
(\ref{coupling2}) and the fact that $\lim_{n \to
\infty}(1-m_n)^n = 1$ for $m_n = o\big(\frac{1}{n}\big)$ (this can
be proved by a simple Taylor series expansion as in \cite[Fact
2]{ZhaoYaganGligor}), the proof of Lemma \ref{lem:cp} is
completed\vspace{2pt} once we demonstrate that with $\text{Var}[X_n]
= o\Big(\frac{\{\mathbb{E}[X_n]\}^2}{ n(\ln n)^2 }\Big)$, there exists
$\epsilon_n = o\big(\frac{1}{\ln n}\big)$ such that
\begin{align}
\bP{X_n <
 (1 - \epsilon_n)\mathbb{E}[X_n]} & = o\bigg(\frac{1}{n}\bigg) , \label{leq1esp_lem}
 \end{align}
 and
 \begin{align}
\bP{X_n >
 (1 + \epsilon_n)\mathbb{E}[X_n]} & = o\bigg(\frac{1}{n}\bigg). \label{geq1esp_lem}
 \end{align}
To prove (\ref{leq1esp_lem}) and (\ref{geq1esp_lem}),
Chebyshev's inequality yields
\begin{align}
\mathbb{P} \big[\hspace{2pt} |X_n-\mathbb{E}[X_n]| >
 \epsilon_n \mathbb{E}[X_n]\big]  & \leq
 \frac{\text{Var}[X_n]}{\big\{\epsilon_n \mathbb{E}[X_n]\big\}^2}.
  \label{thm:grig:Xbound_lem}
 \end{align}
We set $\epsilon_n$ by $\epsilon_n =
\sqrt[4]{\frac{n\text{Var}[X_n]}{\big\{\mathbb{E}[X_n]\big\}^2}} \cdot
\frac{1}{\sqrt{\ln n}} $.
 Then  given condition $\text{Var}[X_n] = o\Big(\frac{\{\mathbb{E}[X_n]\}^2}{ n(\ln
n)^2 }\Big)$, we obtain
\begin{align}
 \epsilon_n & =
 o\Bigg(
  \sqrt[4]{\frac{1}{(\ln n)^2}} \hspace{2pt}\Bigg) \cdot \frac{1}{\sqrt{\ln n}}
   = o\Big(\frac{1}{\ln
n}\Big), \label{eps_lem}
\end{align}
and
\begin{align}
\frac{\text{Var}[X_n]}{\big\{\epsilon_n \mathbb{E}[X_n]\big\}^2}  & =
\sqrt{\frac{\text{Var}[X_n]}{n\big\{\mathbb{E}[X_n]\big\}^2}} \cdot
\ln n
  = o\bigg(\frac{1}{n}\bigg).
\label{varX_lem}
 \end{align}
By (\ref{thm:grig:Xbound_lem}) (\ref{eps_lem}) and (\ref{varX_lem}),
it is straightforward to see that (\ref{leq1esp_lem}) and
(\ref{geq1esp_lem}) hold with $\epsilon_n = o\big(\frac{1}{\ln
n}\big)$. Therefore, we have completed the proof of Lemma
\ref{lem:cp}. \vspace{-1pt} 

\subsection{The Proof of Lemma \ref{cp_rig_er}}  \label{CouplingbinomialER}

In view of \cite[Theorem 1]{zz}, if ${p_n}^2 P_n <   1$ and
$p_n=o\left(\frac{1}{ n}\right)$, with $\hat{p}_n: = {p_n}^2 P_n \cdot \left(1 - n{p_n} + 2 {p_n} -
\frac{{p_n}^2 P_n}{2} \right)$,
then (\ref{cp_res_rig_er}) follows. Given conditions $p_n = O\left(
\frac{1}{n\ln n} \right)$ \vspace{2pt}and ${p_n}^2 P_n =
 O\left( \frac{1}{\ln n} \right)$ in Lemma \ref{cp_rig_er}, ${p_n}^2 P_n <   1$ and
\vspace{2pt}$p_n=o\left(\frac{1}{ n}\right)$ clearly hold. Then
Lemma \ref{cp_rig_er} is proved once we show $\hat{p}_n = {p_n}^2 P_n \cdot
\left[1-
 O\left(\frac{1}{ \ln n}\right)\right]$, which is easy to see via
 \begin{align}
&  - n{p_n} + 2 {p_n} - \frac{{p_n}^2 P_n}{2} \nonumber \\
& \hspace{1pt} = \hspace{-1pt} (-n\hspace{-1pt}+\hspace{-1pt}2)
 \hspace{-1pt}\cdot\hspace{-1pt} O\left(
\frac{1}{n\ln n} \right) \hspace{-1pt}-\hspace{-1pt} \frac{1}{2}
\hspace{-1pt}\cdot\hspace{-1pt} O\left( \frac{1}{\ln n} \right)
\hspace{-1pt}=\hspace{-1pt} -
 O\left(\frac{1}{ \ln n}\right) . \nonumber
\end{align}
Hence, the proof of Lemma \ref{cp_rig_er} is completed. 

\subsection{The Proof of Lemma \ref{cp_urig_rig}} \label{Couplingcp_urig_rig}

We use Lemma \ref{rkgikg} to prove Lemma \ref{cp_urig_rig}. From  \vspace{2pt} $K_n = \omega\left( \ln n \right)$ and $p_n =
\frac{K_n}{P_n}
 \left(1 - \sqrt{\frac{3\ln
n}{K_n }}\hspace{2pt}\right)$,  \vspace{2pt} we first obtain $p_n P_n  = \omega\left( \ln n \right)$ and then
 for all $n$ sufficiently large,
\begin{align}
&  K_n - \left[ p_n P_n + \sqrt{3(p_n P_n + \ln n) \ln n}
\hspace{1.5pt}\right] \nonumber \\ & = K_n \sqrt{\frac{3\ln n}{K_n
}} - \sqrt{3\left[ K_n \left(1 - \sqrt{\frac{3\ln n}{K_n
}}\hspace{2pt}\right) + \ln n\right] \ln n} \nonumber
\\  & = \sqrt{3K_n\ln n}  -
\sqrt{3\left[K_n  \hspace{-1pt}+ \hspace{-1pt} \sqrt{\ln n} \left(
\sqrt{\ln n} \hspace{-1pt}- \hspace{-1pt}
\sqrt{3K_n}\hspace{2pt}\right) \right ] \hspace{-1pt} \ln n}
\nonumber \\  & \geq  0. \nonumber
\end{align}
Then by  Lemma \ref{rkgikg}, Lemma \ref{cp_urig_rig} is now
established. 

\section{Conclusion and Future Work}
\label{sec:Conclusion}

Under a general random intersection graph model, we derive sharp
zero--one laws for $k$-connectivity and $k$-robustness, as well as
the asymptotically exact probability of $k$-connectivity,
 where $k$ is an arbitrary positive integer. A future direction is to obtain the asymptotically exact probability of $k$-robustness for a precise characterization on the robustness strength.

%
%
%

\appendix

\subsection{A lemma to confine $|\alpha_n|$ in Theorems \ref{thm:rig} and \ref{thm:rig:rb} 
 as $o(\ln n)$} \label{seca:conf:bin}

We present Lemma \ref{graph_Hs_cpln} to confine $|\alpha_n|$ in Theorems \ref{thm:rig} and \ref{thm:rig:rb} as $o(\ln n)$; i.e., if Theorems \ref{thm:rig} and \ref{thm:rig:rb} hold under an extra condition $|\alpha_n| = o(\ln n)$, then they also hold regardless of this condition.

\begin{lem} \label{graph_Hs_cpln}

\textbf{(a)} For graph $G_b(n,P_n,p_n)$ under
\begin{align}
   {p_n}^{2}{P_n}   &  = \frac{\ln  n + {(k-1)} \ln \ln n + {\beta_n}}{n}  \label{al0-parta-Hs-od}
\end{align}
with $\lim_{n \to \infty}\beta_n = -\infty$, there exists graph $G_b(n, P_n,\widetilde{p_n})$ under
\begin{align}
  {\widetilde{p_n}}^{2}{P_n}  &  = \frac{\ln  n + {(k-1)} \ln \ln n + {\widetilde{\beta_n}}}{n}  \label{al0-parta-Hs}
\end{align}
with $\lim_{n \to \infty}\widetilde{\beta_n} = -\infty$ and $\widetilde{\beta_n} = -o(\ln n)$
such that $G_b(n,P_n,p_n) \preceq G_b(n,P_n,\widetilde{p_n}) $.


 \textbf{(b)} For graph $G_b(n,P_n,p_n)$ under
\begin{align}
  {p_n}^{2}{P_n}  &  = \frac{\ln  n + {(k-1)} \ln \ln n + {\beta_n}}{n}  \label{al0-parta-Hs-pb-od}
\end{align}
with $\lim_{n \to \infty}\beta_n = \infty$, there exists graph $G_b(n, P_n,\widehat{p_n})$ under
\begin{align}
 {\widehat{p_n}}^{2}{P_n}   &  = \frac{\ln  n + {(k-1)} \ln \ln n + {\widehat{\beta_n}}}{n}  \label{al0-parta-Hs-pb}
\end{align}
with $\lim_{n \to \infty}\widehat{\beta_n} = \infty$ and $\widehat{\beta_n} = o(\ln n)$
such that $G_b(n,P_n,\widehat{p_n}) \preceq G_b(n,P_n,p_n)$.


\end{lem}

The proof of Lemma \ref{graph_Hs_cpln} is given in Section \ref{sec:pro:graph_Hs_cpln}.

We now explain that given Lemma \ref{graph_Hs_cpln}, if Theorems \ref{thm:rig} and \ref{thm:rig:rb} hold under the extra condition
$|\alpha_n| = o(\ln n)$, then they also hold regardless of  the extra condition. Note that results (\ref{bin-kon-e}) and (\ref{bin-kon-e-mnd}) both have a condition $\lim_{n \to \infty}{\alpha_n}
=\alpha ^* \in (-\infty, \infty)$, which clearly implies $|\alpha_n| = o(\ln n)$. Hence, we only need to look at results (\ref{bin-kon-0}) (\ref{bin-kon-1}) (\ref{bin-kon-0-mnd}) (\ref{bin-kon-1-mnd}) (\ref{bin-krb-0})  and (\ref{bin-krb-1}). In particular, we will show that
\begin{align}
\begin{array}{l}\text{if
(\ref{bin-kon-0}) (\ref{bin-kon-0-mnd}) and (\ref{bin-krb-0}) hold under condition
$\alpha_n = -o(\ln n)$,}\\\text{then they also hold regardless of   the  condition.}\end{array}\label{seecpl}
\end{align}
and
\begin{align}
\begin{array}{l}\text{if
 (\ref{bin-kon-1}) (\ref{bin-kon-1-mnd}) and (\ref{bin-krb-1}) hold under condition
$\alpha_n = o(\ln n)$,}\\\text{then they also hold regardless of   the  condition.}\end{array}\label{seecplb}
\end{align}
 To see (\ref{seecpl}), we use Lemma \ref{graph_Hs_cpln}-Property (a) and Lemma \ref{mono-gcp}, and note that $k$-connectivity, the property of minimum node degree being at least $k$, and $k$-robustness are all monotone increasing graph properties. Then with $\mathcal{J}$ denoting any one of the above three properties, for graph $G_b(n,P_n,p_n)$ under (\ref{al0-parta-Hs-od})  with $\lim_{n \to \infty}\beta_n = -\infty$, there exists graph $G_b(n, P_n,\widetilde{p_n})$ under (\ref{al0-parta-Hs}) with $\lim_{n \to \infty}\widetilde{\beta_n} = -\infty$ and $\widetilde{\beta_n} = -o(\ln n)$ such that
 \begin{align}
&  \bP{G_b(n,P_n,p_n) \text{ has $\mathcal{J}$.}} \nonumber \\
&    \leq \bP{G_b(n, P_n,\widetilde{p_n})\text{ has $\mathcal{J}$.}}.
 \label{seecpl1}
\end{align}
If (\ref{bin-kon-0}) and (\ref{bin-krb-0}) hold under  condition
$\alpha_n=- o(\ln n)$, then we use them on graph $G_b(n, P_n,\widetilde{p_n})$ to obtain
 \begin{align}
&\lim_{n \to \infty} \bP{G_b(n, P_n,\widetilde{p_n})\text{ has $\mathcal{J}$.}} = 0,
 \label{seecpl3}
\end{align}
Therefore, (\ref{seecpl1}) and (\ref{seecpl3}) yield
 \begin{align}
&\lim_{n \to \infty} \bP{G_b(n,  {P_n}, {p_n})\text{ has $\mathcal{J}$.}} = 0.
\nonumber
\end{align}
In other words, for graph $G_b(n,  {P_n}, {p_n})$, with $\alpha_n$ in Theorems \ref{thm:rig} and \ref{thm:rig:rb} replaced by $\beta_n$, (\ref{bin-kon-0}) (\ref{bin-kon-0-mnd}) and (\ref{bin-krb-0}) hold. Note that
 we do not have any constraint on whether $\beta_n$ can be expressed as $ -o(\ln n)$. Hence, the arguments above establish   (\ref{seecpl}). The proof of (\ref{seecplb}) using Lemma \ref{graph_Hs_cpln}-Property (b) is similar to that of (\ref{seecpl}) using Lemma \ref{graph_Hs_cpln}-Property (a). We omit the details for simplicity.

\subsection{A lemma to confine $|\alpha_n|$ in Theorems \ref{thm:urig} and \ref{thm:urig:rb} as $o(\ln n)$}  \label{seca:conf:unig}

We present Lemma \ref{graph_Gs_cpl} to confine $|\alpha_n|$ in Theorems \ref{thm:urig} and \ref{thm:urig:rb} as $o(\ln n)$; i.e., if Theorems \ref{thm:urig} and \ref{thm:urig:rb} hold under an extra condition $|\alpha_n| = o(\ln n)$, then they also hold regardless of this condition.

\begin{lem} \label{graph_Gs_cpl}

\textbf{(a)} For graph $G_u(n, P_n, K_n)$ under $P_n = \Omega(n)$ and
\begin{align}
   \frac{{K_n}^{2}}{{P_n}}  &  = \frac{\ln  n + {(k-1)} \ln \ln n + {\beta_n}}{n} \label{al1-parta}
\end{align}
with $\lim_{n \to \infty}\beta_n = -\infty$, there exists graph $G_u(n,P_n, \widetilde{K_n})$ under $P_n = \Omega(n)$ and
\begin{align}
 \frac{{\widetilde{K_n}}^{2}}{{P_n}}  &  = \frac{\ln  n + {(k-1)} \ln \ln n + {\widetilde{\beta_n}}}{n} \label{al0-parta}
\end{align}
with $\lim_{n \to \infty}\widetilde{\beta_n} = -\infty$ and $\widetilde{\beta_n} = -o(\ln n)$,
such that $G_u(n, P_n, K_n) \preceq G_u(n,P_n, \widetilde{K_n})$.

%

\textbf{(b)} For graph $G_u(n, P_n, K_n)$ under $P_n = \Omega(n)$ and
\begin{align}
   \frac{{K_n}^{2}}{{P_n}} &  = \frac{\ln  n + {(k-1)} \ln \ln n + {\beta_n}}{n} \label{al1}
\end{align}
with $\lim_{n \to \infty}\beta_n = \infty$, there exists graph $G_u(n,P_n,\widehat{K_n})$ under $P_n = \Omega(n)$ and
\begin{align}
 \frac{{\widehat{K_n}}^{2}}{{P_n}}  &  = \frac{\ln  n + {(k-1)} \ln \ln n + {\widehat{\beta_n}}}{n} \label{al0}
\end{align}
with $\lim_{n \to \infty}\widehat{\beta_n} = \infty$ and $\widehat{\beta_n} = o(\ln n)$,
such that $G_u(n,P_n,\widehat{K_n})  \preceq G_u(n, P_n, K_n)$.


\end{lem}

The proof of Lemma \ref{graph_Gs_cpl} is given in Section \ref{sec_graph_Gs_cpl}.

We now explain that given Lemma \ref{graph_Gs_cpl}, if Theorems \ref{thm:urig} and \ref{thm:urig:rb} hold under the extra condition
$|\alpha_n| = o(\ln n)$, then they also hold regardless of  the extra condition. Note that results (\ref{uni-kon-e}) and (\ref{uni-kon-e-mnd}) both have a condition $\lim_{n \to \infty}{\alpha_n}
=\alpha ^* \in (-\infty, \infty)$, which clearly implies $|\alpha_n| = o(\ln n)$. Hence, we only need to look at results (\ref{uni-kon-0}) (\ref{uni-kon-1}) (\ref{uni-kon-0-mnd}) (\ref{uni-kon-1-mnd})  (\ref{uni-krb-0})  and (\ref{uni-krb-1}).   In particular, we need to show that
\begin{align}
\begin{array}{l}\text{if
(\ref{uni-kon-0}) (\ref{uni-kon-0-mnd}) and (\ref{uni-krb-0}) hold under condition
$\alpha_n = -o(\ln n)$,}\\\text{then they also hold regardless of   the  condition.}\end{array}\label{seecpluni}
\end{align}
and
\begin{align}
\begin{array}{l}\text{if
 (\ref{uni-kon-1}) (\ref{uni-kon-1-mnd}) and (\ref{uni-krb-1}) hold under condition
$\alpha_n = o(\ln n)$,}\\\text{then they also hold regardless of   the  condition.}\end{array}\label{seecplbuni}
\end{align}
The process of proving (\ref{seecpluni}) and (\ref{seecplbuni}) using Lemma \ref{graph_Gs_cpl} is the same as the above process of  proving (\ref{seecpl}) and (\ref{seecplb}) using Lemma \ref{graph_Hs_cpln}. For brevity, we do not repeat the details here.

\subsection{Proof of Lemma \ref{graph_Hs_cpln}} \label{sec:pro:graph_Hs_cpln}

 \textbf{Proving property (a):}

We set
\begin{align}
\widetilde{\beta_n} = \max\{\beta_n, -\ln \ln n\}. \label{wPntdn2}
\end{align}

Given (\ref{wPntdn2}) and $\lim_{n \to \infty}\beta_n = -\infty$, we  obtain $\lim_{n \to \infty}\widetilde{\beta_n} = -\infty$ and $\widetilde{\beta_n} = -o(\ln n)$. We use $\widetilde{\beta_n} = -o(\ln n)$ and (\ref{al0-parta-Hs}) to have $  {\widetilde{p_n}}^{2}{P_n}  \sim \frac{\ln n}{n}$, so
it is clear for all $n$ sufficiently large that $\widetilde{p_n}$ is less than $1$  and can be used as a probability. Under $p_n \leq \widetilde{p_n}$, by \cite[Section 3]{zz}, there exists a graph coupling under which $G_b(n,P_n,p_n)$ is a spanning subgraph of $G_b(n,P_n,\widetilde{p_n}) $; i.e., $G_b(n,P_n,p_n) \preceq G_b(n,P_n,\widetilde{p_n}) $.

 \textbf{Proving property (b):}

 We set
\begin{align}
\widehat{\beta_n} = \min\{\beta_n, \ln \ln n\}. \label{wPntdn2-pb}
\end{align}

Given (\ref{wPntdn2-pb}) and $\lim_{n \to \infty}\beta_n = \infty$, we clearly obtain $\lim_{n \to \infty}\widehat{\beta_n} = \infty$ and $\widehat{\beta_n} = o(\ln n)$.

It holds from (\ref{wPntdn2-pb}) that $\widehat{\beta_n}  \leq \beta_n$, which along with (\ref{al0-parta-Hs-pb-od})
and (\ref{al0-parta-Hs-pb}) yields $p_n \geq \widehat{p_n}$. Under $p_n \geq \widehat{p_n}$, by \cite[Section 3]{zz}, there exists a graph coupling under which $G_b(n,P_n,p_n)$ is a spanning supergraph of $G_b(n,P_n,\widehat{p_n}) $; i.e., $G_b(n,P_n,\widehat{p_n}) \preceq G_b(n,P_n,p_n)$.

\subsection{The Proof of Lemma \ref{graph_Gs_cpl}} \label{sec_graph_Gs_cpl}

 \textbf{Proving property (a):}

 We
 define $\widetilde{\beta_n}^*$ by
 \begin{align}
\widetilde{\beta_n}^* &  = \max\{\beta_n, -\ln \ln n\}, \label{al2-parta}
\end{align}
and
define $\widetilde{K_n}^*$ such that
\begin{align}
   \frac{({\widetilde{K_n}^{*}})^{2}}{{{P_n}}}  &  = \frac{\ln  n + {(k-1)} \ln \ln n + \widetilde{\beta_n}^*}{n}. \label{al3-parta}
\end{align}
Note that $\widetilde{K_n}^*$ might or might not be an integer.
We set
\begin{align}
\widetilde{K_n} & : = \big\lfloor \widetilde{K_n}^* \big\rfloor,  \label{al4-parta}
\end{align}
where the floor function $\lfloor x \rfloor$ means the largest integer not greater than $x$.

From (\ref{al1}) (\ref{al2-parta}) and (\ref{al3-parta}), it holds that
\begin{align}
K_n \leq \widetilde{K_n}^*.  \label{Kn1-parta}
\end{align}
Then by (\ref{al4-parta}) (\ref{Kn1-parta}) and the fact that $K_n$ and $\widetilde{K_n}$ are both integers, it follows that
\begin{align}
K_n \leq \widetilde{K_n}.  \label{al6-parta}
\end{align}
 From (\ref{al6-parta}), by  \cite[Lemma 3]{Rybarczyk}, there exists a graph coupling under which $G_u(n,P_n,K_n)$ is a spanning subgraph of $G_u(n,P_n,\widetilde{K_n})$; i.e., $G_u(n, P_n, K_n) \preceq G_u(n,P_n, \widetilde{K_n})$. Therefore, the proof of property (a) is completed once we show
$\widetilde{\beta_n}$ defined in $(\ref{al0-parta})$ satisfies
\begin{align}
  \lim_{n \to \infty}\widetilde{\beta_n} & = - \infty, \label{al8-parta} \\
\text{and }\widetilde{\beta_n} & = - o(\ln n).  \label{al7-parta}
\end{align}

We first prove (\ref{al8-parta}). From (\ref{al0-parta}) (\ref{al3-parta}) and (\ref{al4-parta}), it holds that
\begin{align}
\widetilde{\beta_n} \leq \widetilde{\beta_n}^*, \label{haa-parta}
\end{align}
which together with (\ref{al2-parta}) and $\lim_{n \to \infty}\beta_n = -\infty$ yields (\ref{al8-parta}).

Now we establish (\ref{al7-parta}). From (\ref{al4-parta}), we have $\widetilde{K_n} > \widetilde{K_n}^* - 1$. Then from (\ref{al0-parta}), it holds that
\begin{align}
 \widetilde{\beta_n} & = n \cdot   \frac{{\widetilde{K_n}}^{2}}{{{P_n}}}  - [\ln  n + {(k-1)} \ln \ln n] \nonumber \\
 & > n \cdot  \frac{{(\widetilde{K_n}^*)^2 - 2\widetilde{K_n}^*}}{{{P_n}}}  - [\ln  n + {(k-1)} \ln \ln n]
  . \label{aph1-parta}
\end{align}
By $\lim_{n \to \infty}\beta_n =- \infty$, it holds that $\beta_n \leq 0$ for all $n$ sufficiently large. Then from (\ref{al2-parta}), it follows that
\begin{align}
 \widetilde{\beta_n}^* = - O(\ln \ln n),  \label{widetilde-al2-parta}
\end{align}
which along with (\ref{al3-parta}) yields
\begin{align}
 \frac{\widetilde{K_n}^*}{P_n} &  \sim \sqrt{ \frac{ \ln n }{nP_n}} = O\bigg(\frac{\sqrt{\ln n}}{n}\bigg) .\label{aph5-parta}
\end{align}

%
%
Applying (\ref{al3-parta}) (\ref{aph5-parta}) and $P_n = \Omega(n)$ to (\ref{aph1-parta}), we obtain
\begin{align}
 \widetilde{\beta_n}  & > \bigg\{n \cdot  \frac{({\widetilde{K_n}^{*}})^{2}}{{{P_n}}}  - [\ln  n + {(k-1)} \ln \ln n] \bigg\} - 2n \cdot  \frac{\widetilde{K_n}^*}{P_n}  \nonumber \\
 &    = \widetilde{\beta_n}^* - O\big(\sqrt{\ln n}\hspace{2pt}\big). \label{widetilde-al-parta}
\end{align}
Thus, from (\ref{haa-parta})  (\ref{widetilde-al2-parta}) and (\ref{widetilde-al-parta}), clearly
$ \widetilde{\beta_n} $ can be written as $- O\big(\sqrt{\ln n}\hspace{2pt}\big)$ and further $- o(\ln n)$; i.e., (\ref{al7-parta}) is proved. Then as explained above, since we have shown (\ref{al8-parta}) and (\ref{al7-parta}), property (a) of Lemma \ref{graph_Gs_cpl} is established.

 \textbf{Proving property (b):}

 We
 define $\widehat{\beta_n}^*$ by
 \begin{align}
\widehat{\beta_n}^* &  = \min\{\beta_n, \ln \ln n\}, \label{al2}
\end{align}
and
define $\widehat{K_n}^*$ such that
\begin{align}
 \frac{({\widehat{K_n}^{*}})^{2}}{{{P_n}}}  &  = \frac{\ln  n + {(k-1)} \ln \ln n + \widehat{\beta_n}^*}{n}. \label{al3}
\end{align}
We set
\begin{align}
\widehat{K_n} & : = \big\lceil \widehat{K_n}^* \big\rceil.  \label{al4}
\end{align}

From (\ref{al1}) (\ref{al2}) and (\ref{al3}), it holds that
\begin{align}
K_n \geq \widehat{K_n}^*.  \label{Kn1}
\end{align}
Then by (\ref{al4}) (\ref{Kn1}) and the fact that $K_n$ and $\widehat{K_n}$ are both integers, it follows that
\begin{align}
K_n \geq \widehat{K_n}.  \label{al6}
\end{align}
 From (\ref{al6}), by  \cite[Lemma 3]{Rybarczyk}, there exists a graph coupling under which $G_u(n,P_n,K_n)$ is a spanning supergraph of $G_u(n,P_n,\widehat{K_n})$; i.e., $G_u(n,P_n,\widehat{K_n})  \preceq G_u(n, P_n, K_n)$. Therefore, the proof of property (b) is completed once we show
$\widehat{\beta_n}$ defined in $(\ref{al0})$ satisfies
\begin{align}
  \lim_{n \to \infty}\widehat{\beta_n} & = \infty, \label{al8} \\
 \text{and }\widehat{\beta_n} & = o(\ln n).  \label{al7}
\end{align}

We first prove (\ref{al8}). From (\ref{al0}) (\ref{al3}) and (\ref{al4}), it holds that
\begin{align}
\widehat{\beta_n} \geq \widehat{\beta_n}^*, \label{haa}
\end{align}
which together with (\ref{al2}) and $\lim_{n \to \infty}\beta_n = \infty$ yields (\ref{al8}).

Now we establish (\ref{al7}). From (\ref{al4}), we have $\widehat{K_n} < \widehat{K_n}^* + 1$. Then from (\ref{al0}), it holds that
\begin{align}
 \widehat{\beta_n} & = n   \cdot \frac{{\widehat{K_n}}^{2}}{{{P_n}}}  - [\ln  n + {(k-1)} \ln \ln n] \nonumber \\
 & \leq  n \cdot   \frac{({\widehat{K_n}^{*}})^{2} + 3  {\widehat{K_n}^{*}} }{{{P_n}}}  - [\ln  n + {(k-1)} \ln \ln n]  . \label{aph1}
\end{align}
By $\lim_{n \to \infty}\beta_n = \infty$, it holds that $\beta_n \geq 0$ for all $n$ sufficiently large. Then from (\ref{al2}), it follows that
\begin{align}
 \widehat{\beta_n}^* = O(\ln \ln n),  \label{widehat-al2}
\end{align}
which along with (\ref{al3}) and condition $P_n = \Omega(n)$ induces
\begin{align}
\frac{\widehat{K_n}^*}{P_n} &  \sim \sqrt{ \frac{ \ln n }{nP_n}} = O\bigg(\frac{\sqrt{\ln n}}{n}\bigg) . \label{aph5}
\end{align}
Hence, we have $\lim_{n \to \infty} \widehat{K_n}^* = \infty$ and it further holds for all $n$ sufficient large that
\begin{align}
{(\widehat{K_n}^* + 1)}^{2}< ({\widehat{K_n}^{*}})^{2} + 3  {\widehat{K_n}^{*}} . \label{aph2}
\end{align}
Applying (\ref{al3}) (\ref{aph5}) and $P_n = \Omega(n)$ to (\ref{aph1}), we obtain
\begin{align}
 \widehat{\beta_n}  & <   \bigg\{n \cdot  \frac{({\widehat{K_n}^{*}})^{2}}{{{P_n}}}  - [\ln  n + {(k-1)} \ln \ln n] \bigg\} + 3 n \cdot  \frac{\widehat{K_n}^*}{P_n}
  \nonumber \\
 &    = \widehat{\beta_n}^* + O\big(\sqrt{\ln n}\hspace{2pt}\big).\label{widehat-al}
\end{align}
Thus, from (\ref{haa})  (\ref{widehat-al2}) and (\ref{widehat-al}), clearly
$ \widehat{\beta_n} $ can be written as $ O\big(\sqrt{\ln n}\hspace{2pt}\big)$ and further $ o(\ln n)$; i.e., (\ref{al7}) is proved. Then as explained above, since we have shown (\ref{al8}) and (\ref{al7}), property (b) of Lemma \ref{graph_Gs_cpl} is established.

%

\end{document}